\documentclass[%
 10pt,
 reprint,
superscriptaddress,
 amsmath,amssymb,
 aps,
pra,
]{revtex4-2}

\usepackage{graphicx}
\usepackage{bm}
\usepackage[colorlinks=true, allcolors=black]{hyperref} 
\usepackage{mathtools, amsmath}
\usepackage[italic=true]{derivative}
\usepackage{braket}

\begin{document}

\preprint{APS/123-QED}

\title{
Vortex Mass in Superfluid Fermi Gases along the BEC--BCS Crossover
}%

\author{Lucas Levrouw}

\email{lucas.levrouw@uantwerpen.be}
\affiliation{%
 Theory of Quantum systems and Complex systems (TQC), University of Antwerp \\
 Universiteitsplein 1, 2610 Antwerpen, Belgium
}%

\author{Hiromitsu Takeuchi}

\email{takeuchi@omu.ac.jp}

\affiliation{
  Nambu Yoichiro Institute of Theoretical and Experimental Physics (NITEP),
  Osaka Metropolitan University, 3-3-138 Sugimoto, Osaka 558-8585, Japan
}%
\affiliation{
  Department of Physics,
  Osaka Metropolitan University, 3-3-138 Sugimoto, Osaka 558-8585, Japan
}%

\author{Jacques Tempere}%
 \email{jacques.tempere@uantwerpen.be}
\affiliation{%
 University of Antwerp, Theory of Quantum systems and Complex systems \\
 Universiteitsplein 1, 2610 Antwerpen, Belgium
}%

\date{\today}

\begin{abstract}
Vortex mass is a key concept in the study of superfluid dynamics, referring to the inertia of vortices in a superfluid, which affects their motion and behavior. Despite being an important quantity, the vortex mass has never been observed experimentally, and remains an unresolved issue in this field. As of now, a large body of research assumes that the vortex mass is a local parameter. In contrast, we present a calculation that suggests a logarithmic dependence on the system size, agreeing with some earlier predictions in the context of Bose gases.  We analyze the problem using an effective field theory that describes ultracold atomic Fermi gases over the BEC--BCS crossover at both zero and nonzero temperatures. Our study reveals a strong dependence of the vortex mass on the scattering length; in particular, the vortex mass grows rapidly when moving toward the BCS side. Furthermore, we find that the system-size dependence of the vortex mass results in values an order of magnitude larger than those predicted by other models for realistic system sizes. This implies that the vortex mass could be observable in a wider parameter range than was previously expected. This is particularly relevant considering recent advances in experimental techniques that place the observation of vortex mass in superfluid Fermi gases within reach.
\end{abstract}

\maketitle


\section{Introduction}

Quantized vortices are among the intriguing phenomena associated with superfluidity, playing a crucial role in superfluid helium, ultracold atomic gases, as well as superconductors.
 Their dynamics are generally complex, but can often be approximated using a point-vortex model, where vortices are treated as point-like particles moving within the velocity field induced by other vortices. Within this framework, one can assign an effective inertia to vortices, known as the vortex mass.
 The vortex mass problem refers to the long-standing challenge of defining and measuring this mass. While various theoretical models have predicted a nonzero vortex mass \citep{popov1973, duan1992, duan1994, baym1983, kopnin1978, kopnin1998}, experimental studies of vortex dynamics in superfluid helium and atomic Bose-Einstein condensates (BECs) have successfully used point-vortex models that assume zero mass \citep{donnelly1991, navarro2013, samson2016}.
In superconductors, a vortex mass has also been predicted \citep{suhl1965}  and has been observed to be nonzero in multiple experiments \citep{fil2007, golubchik2012, tesar2021, nakamura2024}. However, also here there is still some controversy, as the reported values vary significantly across different systems and measurement methods.
The difficulty in measuring vortex mass in helium can be understood in terms of its core size, which is significantly smaller than other characteristic length scales in the system. As a result, any inertial effects are expected to be negligible. In contrast, in ultracold gases, where the vortex core size can be comparable to other relevant length scales, the vortex mass may play a significant role in the system’s dynamics.
Experiments creating a solitonic vortex in a harmonically trapped cigar-shaped cloud \cite{yefsah2013, ku2014} report a nonzero vortex mass; however the geometry makes it difficult to compare to theoretical estimates for vortices moving in a plane. Recent developments in experiments on ultracold Fermi gases \citep{kwon2021, delpace2022,hernandez-rajkov2024, grani2025} seem to place the measurement of the vortex mass in a planar superfluid system within experimental reach. In these experiments, a quasi-two-dimensional condensate is created in a box trap, and individual vortices can be generated and manipulated, enabling the exploration of virtually any initial conditions to study vortex dynamics.
In addition, the vortex mass is predicted to be significant in two-component BECs \citep{richaud2020,richaud2021,bellettini2023}. Recent experiments have shown excellent control over interspecies interactions and the possibility of species-selective trapping potentials \citep{wacker2015, franzen2022}.
An experimental setup is currently being developed that combines such a highly tunable two-component Bose gas with vortex creation and manipulation techniques previously demonstrated in single-component BECs \citep{wilson2022, wilson2024APS}.

We should emphasize that even defining the vortex mass requires us to only consider situations in which a point-vortex model is applicable. More correct would be to use a field theory for the order parameter. The point-vortex model only includes pairwise interactions, and assumes that the vortices are moving sufficiently slowly and that the distance between the vortices is large enough compared to the healing length. 
Nevertheless, the point-vortex model is a useful tool to simulate more complicated vortex dynamics. Moreover, it is easier to test, since it only relies on a few parameters, and leads to a clearer interpretation of the dynamics.
How exactly to define and calculate the vortex mass is still unclear.
From the theoretical side, there have been multiple proposals to calculate the vortex mass.
\citet{popov1973} proposed a mass $
     M_r = \mathcal E / c^2 
$ based on a relativistic analogy,
where
$\mathcal E$ is the energy of a straight vortex (per unit length), and $c$ the speed of sound. 
 The same expression was obtained by \citet{duan1992,duan1994} with a different argument, based on the compressibility of the fluid.
It is important to note that, as the energy of a vortex is logarithmically dependent on the system size, also the relativistic vortex mass has this dependence. This suggests that like the energy, the vortex mass is a global feature of the system---we will refer to this as a \emph{global} vortex mass. We note that global vortex masses lead to the additional complication that the mass may depend on the position of the vortex relative to the system boundary; in this case, the vortex should additionally be assumed to be close enough to the center of the system.
In contrast, other proposals consider masses that are localized within the vortex core and can therefore be considered \emph{local} vortex masses.
By approximating the vortex as having a cylindrical rigid core, \citet{baym1983} obtained a contribution to the local vortex mass, corresponding to the associated (or induced) mass in hydrodynamics, which was found to be equal to the mass of the superfluid expelled by the vortex core. Another contribution to the local vortex mass comes quite simply from matter being present in the core. For mixtures of Bose gases, where the minority component accumulates in the vortex core, this is predicted to lead to a vortex mass \citep{richaud2020,richaud2021,bellettini2023}.
For single-component gases, it has been argued that bound quasiparticles in the vortex core also give rise to a vortex mass, both in Fermi gases [the so-called Caroli–de Gennes–Matricon (CdGM) states, leading to the Kopnin mass] \citep{kopnin1978, kopnin1998} and in Bose gases \citep{simula2018}.

In this paper, we will compute the local and global vortex mass in superfluid Fermi gases based on an effective field theory that describes the BEC--BCS crossover \citep{klimin2015, klimin2014}.
In Sec. \ref{sec:theory} we introduce the theoretical model we will use to calculate the vortex mass. First, we give a definition of the vortex mass in the context of the two-fluid model. Next, we introduce the effective field theory to describe fermions across the BEC--BCS crossover and calculate superfluid and normal densities.
Section \ref{sec:vortex-profile} reveals the radial profile of the superfluid and normal densities around a quantum vortex at various interaction strengths.
In Sec. \ref{sec:vortex-mass} we present our main result. We propose an analytical model to describe the vortex mass and its dependence on the system size. We characterize the vortex mass along the BEC--BCS crossover and consider finite-temperature effects. We discuss how these results affect the observability of the vortex mass.

\section{Vortex mass in the effective field theory framework} \label{sec:theory}

The point-vortex model \footnote{We briefly review the point-vortex model with vortex inertia in Appendix \ref{app:pvm}}, in which the vortex mass is defined, is based on the assumptions that the vortices are point-like objects and that the velocity field induced by a vortex is circularly symmetric.
In addition to assuming the rotational symmetry, we will suppose that the vortex maintains its equilibrium profile, allowing all properties to be determined from the stationary state. Our model will be introduced in two steps: first, by defining expressions for the vortex mass within the two-fluid model of superfluidity, and then by applying them to Fermi gases using an effective field theory.

\subsection{Associated and internal masses}

In classical hydrodynamics, it is well established that a rotating cylinder in an irrotational fluid acquires an effective inertia of consisting of the mass fluid expelled from the volume of the cylinder (called the associated mass) on top of the mass of the cylinder (which we call the internal mass) \citep{lamb1997}.
According to the two-fluid model, a quantum fluid below the superfluid transition temperature is considered to consist of superfluid and normal components with mass densities $\rho_s$ and $\rho_n$, respectively.
To the former, we will attribute an associated mass, analogous to that in hydrodynamics. To the latter, we will attribute an internal mass, corresponding to the non-superfluid matter accumulated in the vortex core. As is the case in classical hydrodynamics, we consider these vortex masses to be additive. In previous literature, they were also treated as independent contributions \citep{volovik1998}.

The superfluid density $\rho_s$ vanishes inside the core of a quantum vortex, the size of which is characterized by the healing length $\xi$.
Naively, the vortex can be approximated by an empty cylinder with a radius of the order $\xi$; in hydrodynamics this leads to the local associated mass of the order $\pi \xi^2 \rho_{s,\infty}$ \citep{baym1983}, equal to the mass of the superfluid removed from the cylindrical vortex core.
Here $\rho_{s,\infty}$ is the superfluid density in the bulk.
In this paper, we propose to generalize this to the reduction in superfluid density caused by the presence of a vortex, which we will call the global associated mass
\begin{equation}
    M_{a} = 2\pi \int_0^\infty dr \, r\left(\rho_{s,\infty} - \rho_s(r)\right)  \label{eq:associated-mass},
\end{equation}
fully taking into account the radial superfluid density profile, where $r$ is the distance from the vortex axis.

Similarly for the internal mass, we do not limit ourselves to only the normal component matter close to the vortex core. Instead, we calculate all the mass accumulated in the core
\begin{equation}
M_{i} = 2\pi \int_0^\infty dr \, r\left(\rho_n(r) - \rho_{n,\infty}\right).\label{eq:internal-mass}
\end{equation}
We refer to Eq. \eqref{eq:internal-mass} as the global internal mass, in contrast to previous results that give a local internal mass like the core-mode calculation of \textcite{kopnin1978}.
We note that at nonzero temperatures, there may also be some nonzero normal density $\rho_{n,\infty}$ in the bulk. We assume that this bulk normal component remains static and thus that the only contribution to the internal mass comes from the normal density in excess of this value. The same procedure was followed by \textcite{richaud2024}.
Let us clarify the scope of application of this model. First, we should note that to even define the vortex mass, we should limit ourselves to the cases in which a point-vortex model is a good approximation, which were discussed in the Introduction. Furthermore, the applicability of the formulas \eqref{eq:associated-mass} and \eqref{eq:internal-mass} fundamentally rests on the assertion that the two-fluid description is valid. Also, the model is not complete without an accurate calculation of the normal and superfluid densities; in the next sections, we will discuss how to do this in the effective field theory framework.
We do not provide a microscopic model to justify Eqs. \eqref{eq:associated-mass} and \eqref{eq:internal-mass}. However, the advantage of our model is that it yields a straightforward evaluation of the vortex mass, still including the full spatial structure of the vortex, which was disregarded in previous literature. Since no experimental observations have been made so far, our theoretical model should be considered as one possibility.

\subsection{Effective field theory}

In order to compute the quantities $M_a$ and $M_i$, we need a way to calculate the
radial profiles $\rho_s(r)$ and $\rho_n(r)$ as a function of distance $r$ from the vortex axis.
The main theoretical tool we will  use for this in this paper is an effective field theory (EFT) for superfluid Fermi gases \citep{klimin2015, vanloon2018}.
The (imaginary-time) action for the EFT is given by
\begin{multline} \label{eq:eft-action}
    S_{EFT}[\Phi,\bar\Phi] =\\
    \int_0^{\hbar \beta} d\tau \int d \mathbf{x} \left[\hbar\frac{D(|\Phi|^2)}{2}\left(\bar \Phi\frac{\partial\Phi}{\partial\tau}- \frac{\partial\bar \Phi}{\partial\tau}\Phi\right) + \hbar^2 Q \frac{\partial \bar \Phi}{\partial \tau}  \frac{\partial \Phi}{\partial \tau} \right.\\
    - \hbar^2 R\left(\frac{\partial{|\Phi|^2}}{\partial \tau}\right)^2 
+ \frac{\hbar^2C}{2m} \left(\nabla \bar \Phi\cdot \nabla \Phi\right)\\ \left.
- \frac{\hbar^2 E}{2m} (\nabla |\Phi|^2)^2
+ \Omega_{sp}(|\Phi|^2)
\right].
\end{multline}
Here, $\Phi = |\Phi| \exp(i\theta)$ is the complex Bardeen--Cooper--Schrieffer (BCS) order parameter. The coefficients $D(|\Phi|^2)$ and $\Omega_{sp}(|\Phi|^2)$ are functions of the order parameter. The coefficients $C$, $E$, $Q$, and $R$ are kept constant and evaluated in the bulk. All of these can be calculated as a function of the $s$-wave scattering length $a_s$ and the temperature $T$, using the expressions found in Appendix \ref{app:eft}, which also outlines the derivation of the EFT.

This can be understood as the Gross--Pitaevskii (classical field) action \cite{pethick2006, pitaevskii2003} with extra gradient corrections. Indeed, the terms with coefficients $\tilde D(|\Phi|^2)$ and $C$ correspond to the usual derivatives occurring in the Gross--Pitaevskii theory, while $\Omega_{sp}(|\Phi|^2)$ contains the nonlinear interaction, as well as higher-order corrections.
The remaining terms have no analog in the Gross--Pitaevskii equation; however, it can be shown that in the BEC limit and at zero temperature, they vanish.
Similarly, the EFT reduces to the Ginzburg--Landau theory when approaching the critical temperature.
The EFT has been successfully applied to solitons and vortices \citep{klimin2014,klimin2016, lombardi2016, vanalphen2024}. An advantage of this method over alternatives like Bogoliubov--de Gennes (BdG) theory is that it is much less computationally expensive.

The EFT was derived based on a gradient expansion, which assumes that the fermionic degrees of freedom, with a characteristic length scale of the pair correlation length, should vary on a shorter scale than the bosonic degrees of freedom, characterized by the healing length \citep{lombardi2016}. Hence, the range of validity of the EFT is limited to the cases where the pair correlation length is small with respect to the healing length. 
According to the BCS theory,  the pair correlation length and the healing length are of similar magnitude at low temperatures and increase with decreasing energy gap $\Delta$ as $\propto 1/\Delta$ and thus diverge as $1/\Delta\propto \exp((k_F|a_s|)^{-1})$ in the BCS limit $(k_Fa_s)^{-1} \to -\infty$ where $k_F$ is the Fermi wave vector \citep{marini1998}. Conversely, when approaching the BEC limit $(k_Fa_s)^{-1} \to +\infty$, the healing length diverges as $(k_F a_s)^{-1/2}$ while the pair correlation length goes to zero.
The healing length also diverges just below the superfluid critical temperature $T_c$ as $\propto (1-T/T_c)^{-1/2}$ while the correlation length remains finite \citep{palestini2014}.
This means that the effective field theory will work best if the temperature is close to the critical temperature, or in the BEC regime. In the relevant experiments in superfluid Fermi gases \citep{kwon2021, delpace2022, hernandez-rajkov2024,grani2025}, the parameters take the values $(k_F a_s)^{-1} \gtrsim -0.5$ and $T/T_c \gtrsim 0.4$, where $k_F$ is the Fermi wave vector and $T_c$ the critical temperature.
In this parameter range, we can still expect good agreement.

We note that the EFT action needs to be complemented by an equation of state $\rho_{\text{tot}}(T, \mu, \Delta)$, which relates the (total) density to the temperature $T$, chemical potential $\mu$ and order parameter $\Delta$. In this paper, we use the mean-field equation of state. Choosing a different equation of state, for example in the Gaussian pair fluctuation approximation \citep{nozieres1985,hu2006}, would allow us to add beyond-mean-field corrections. These corrections are particularly important in the BEC regime, and at high temperatures. We will comment further on the limitations of this approximation in the discussion of the results.

\subsection{Superfluid and normal densities}

In addition to being significantly easier to solve than the BdG equations, the EFT also allows for straightforward calculation of physical quantities of interest. Here, we will derive expressions for the superfluid and normal densities as a function of the order parameter $\Phi$.

For the total (mass) density, we can use the local-density approximation (LDA) of the mean-field density function \cite{klimin2015}. This is given by
\begin{equation}
\rho_{\text{tot}}(|\Phi|^2)
= m \int \frac{d\mathbf k}{(2 \pi)^3}\left[1-\frac{\xi_{\mathbf{k}}}{E_{\mathbf{k}}(|\Phi|^2)} X(E_{\mathbf k}(|\Phi|^2)) \right], \label{eq:total-density}
\end{equation}
where $\xi_{\mathbf k} = k^2/2m-\mu$ and $E_{\mathbf k}(|\Phi|^2) = \sqrt{\xi_k^2 + |\Phi|^2}$ are the single-particle dispersion and Bogoliubov dispersion respectively, and $X(\epsilon) = \tanh(\beta\epsilon/2)$.

The superfluid density $\rho_s$ can be identified as the phase stiffness, which is defined so that the term in the Lagrangian proportional to the $(\nabla\theta)^2$ takes the form
\begin{equation}
\frac{\hbar^2C}{2m} |\Phi|^2 (\nabla\theta)^2
= \frac{1}{2} \rho_s \mathbf v_s^2,
\end{equation}
where the superfluid velocity $\mathbf v_s$ is defined as $\mathbf v_s = \hbar \nabla\theta/2m$ for Cooper pairs with mass $2m$ \citep{giorgini2008}.
From the action \eqref{eq:eft-action}, we find \footnote{The same expression appeared before in Ref. \citep{klimin2015}, but with a wrong prefactor.}
\begin{equation} \label{eq:superfluid-density-eft}
 \rho_s(|\Phi|^2) = 4mC |\Phi|^2.
\end{equation}
Then, we may define the normal density as the difference between total and superfluid densities
\begin{equation} \label{eq:normal-density-eft}
\rho_n(|\Phi|^2) = \rho_{\text{tot}}(|\Phi|^2) - \rho_s(|\Phi|^2).
\end{equation}
In the bulk, the order parameter can be calculated using the mean-field equations for the spatially uniform case \citep{tempere2012}.
Writing $\Delta$ for the bulk value of the order parameter, the normal density can be written as
\begin{equation}
    \rho_{n,\infty} 
    = \frac{\hbar^2}{3} \int \frac{d\mathbf k}{(2\pi)^3} \,k^2 Y(E_{\mathbf k}(\Delta)),
\end{equation}
where $Y(\epsilon) = \partial X / \partial \epsilon = \beta/(\cosh(\beta\epsilon)+1)$
is the the Yoshida distribution.
Accordingly, the bulk superfluid density is computed as  $\rho_{s,\infty}=\rho_{\infty}-\rho_{n,\infty}$ with $\rho_{\infty} = \rho_{\text{tot}}(\Delta^2)$. These findings are consistent with mean-field results 
in the literature for the uniform case \citep{lifshitz1980, botelho2006, tempere2009}. The mean-field density function defined in \eqref{eq:total-density} does not include contributions corresponding to fluctuations of the order parameter. \textcite{taylor2006} argued that these do not modify the superfluid density, but only introduce an extra contribution to the normal density.

\section{Vortex profile} \label{sec:vortex-profile}
\subsection{Asymptotic form of the density}

In the previous section, expressions for the superfluid and normal densities as a function of the order parameter were obtained. To find the radial dependence of these quantities,
we need to find a solution of the EFT equation of motion, which in the stationary case reduces to
\begin{equation}
    -\frac{\hbar^2C}{2m} \nabla^2\Phi+\left( \mathcal A(|\Phi|^2) 
 + \frac{\hbar^2E}{m} \nabla^2 |\Phi|^2\right) \Phi= 0
\end{equation}
where
\begin{align}
\mathcal{A}(|\Phi|^2)=\frac{\partial \Omega_{sp}}{\partial|\Phi|^2}.
\end{align}
 For a rotationally symmetric, singly quantized vortex positioned in the origin, the order parameter should take the following form in polar coordinates $(r, \varphi)$
\begin{equation} \label{eq:vortex-ansatz}
    \Phi(r, \varphi) = \Delta \, f(r) e^{i \varphi},
\end{equation}
where as before, $\Delta$ is the value of the order parameter in the bulk.
The order parameter profile $f(r)$ satisfies the following equation
\begin{multline}
\label{eq:eft-profile-equation}
-\frac{\hbar^2C}{2m} \left(\frac{1}{r}\frac{d}{dr}[r f'(r)] - \frac{1}{r^2} f(r)\right) \\
+ \Delta^2\frac{\hbar^2E}{m} \frac{f(r)}{r} \frac{d}{dr}\left[r \frac{d}{dr} (f(r)^2)\right]\\
+ \mathcal A( \Delta^2 f(r)^2) f(r) = 0.
\end{multline}

\begin{figure}
\includegraphics[width=\columnwidth]{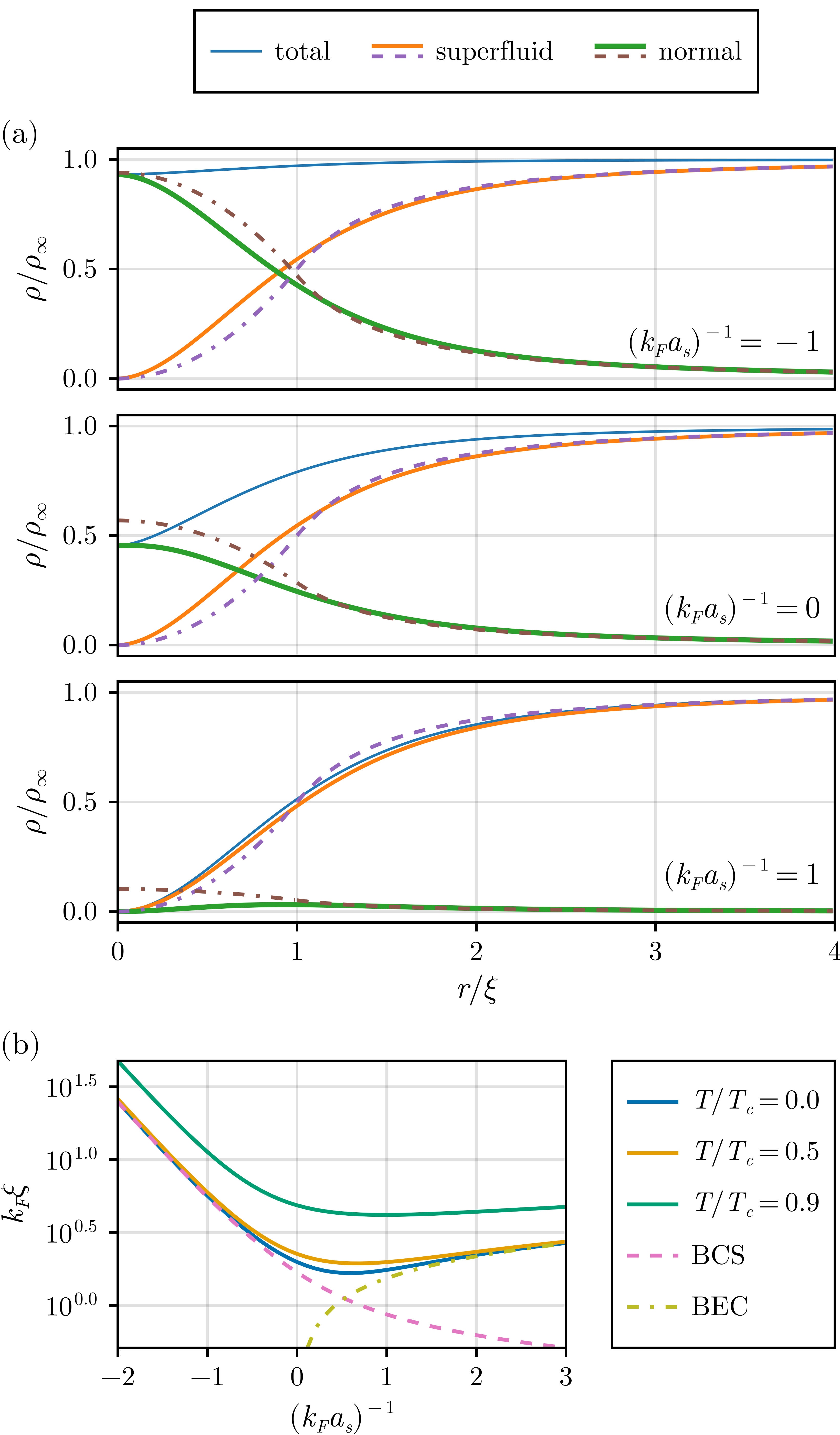}
    \caption{
    (a) Total density $\rho_{\text{tot}}$ (solid blue line), superfluid density $\rho_s$ (solid orange line) and normal density $\rho_n$ (solid green line) as a function of the radial coordinate, calculated in the EFT framework at temperature zero. The densities are normalized by the total density at $r=\infty$.
    Also plotted are the asymptotic expressions for the superfluid (dashed purple line) and normal (dashed brown line) densities. In the $r < \xi$ region, the asymptotics are continued by dash-dotted lines indicating the simplified analytic vortex profile introduced in Eqs. \eqref{eq:simplified-rhos} and \eqref{eq:simplified-rhon}.
    (b) Healing length as defined in Eq. \eqref{eq:eft-xi1} over the crossover, for various temperatures. These are compared to analytic expressions in the BCS and BEC limits at zero temperature, which are derived in Appendix \ref{app:healing}.}
    \label{fig:profiles-densities-T0}
\end{figure}

In the EFT formalism, it is straightforward to check how the profile should behave asymptotically.
Expanding Eq. \eqref{eq:eft-profile-equation} around $f(r) = 1$ leads to the following asymptotics
\begin{equation} \label{eq:profile-asymptotic}
    f(r) \sim 1 -\frac{1}{4}\frac{\xi^2}{r^2},
\end{equation}
where
\begin{equation} \label{eq:eft-xi1}
 \xi = \sqrt{\frac{\hbar^2}{m}\frac{C}{\Delta^2G}},
\quad G = \left.\pdv{\mathcal A}{|\Phi|^2}\right|_\infty,
\end{equation}
writing a bar with subscript $\infty$ to denote evaluation of the derivative at $|\Phi|=
\Delta$, corresponding to $r = \infty$. This quantity is plotted in Fig. \ref{fig:profiles-densities-T0}(b).
 In the BEC limit at zero temperature, $\xi$ tends to the bosonic healing length, so that we recover the asymptotic behavior of the Gross--Pitaevskii theory and the Ginzburg-Landau theory \citep{pitaevskii1961,chen1994, xu1995}.
 \citet{simonucci2013} note that profiles calculated in the Bogoliubov--de Gennes framework also follow the same characteristic decay. However, outside the validity region of the EFT, the value of $\xi$ is altered from the value in Eq. \eqref{eq:eft-xi1}. We note that the hyperbolic tangent, which is a commonly used Ansatz for the order parameter of a vortex \citep{verhelst2017, berthod2016, ichmoukhamedov2020, vanalphen2024}, cannot reproduce this asymptotic form.
 
 Next, we consider the asymptotic behavior of the superfluid and normal densities, based on expressions \eqref{eq:superfluid-density-eft} and \eqref{eq:normal-density-eft}.
Using
\begin{equation}
    1 - f(r)^2 \sim \frac{1}{2}\frac{\xi^2}{r^2},
\end{equation}
the superfluid and normal densities can be expanded around their bulk value as
\begin{align}
    \rho_s &= \rho_{s,\infty} \left(1 -\frac{1}{2}\frac{\xi^2}{r^2}\right) + O\left(\frac{\xi^4}{r^4}\right) \label{eq:asymptotic-superfluid} \\
    \rho_n &= \rho_{n,\infty} + \delta\rho_{n,\infty}\frac{1}{2}\frac{\xi^2}{r^2} + O\left(\frac{\xi^4}{r^4}\right). \label{eq:asymptotic-normal}
\end{align}
Here we introduced 
\begin{equation}
    \delta\rho_{n,\infty} = - \Delta^2  \left.\pdv{\rho_n}{ |\Phi|^2}\right|_\infty
    =
    \rho_{s,\infty} - \Delta^2   \left.\pdv{\rho}{ |\Phi|^2}\right|_\infty.
\end{equation}
An explicit expression is given in Appendix \ref{app:eft}.
In Fig. \ref{fig:profiles-densities-T0}, these asymptotics are plotted together with the numerically computed density profiles.
 
\subsection{Numerical results}

 We solve Eq. \eqref{eq:eft-profile-equation} using an imaginary time evolution method, which is described in Appendix \ref{app:num}. In previous work, profiles of the order parameter and total density were already compared to the Bogliubov-de Gennes theory \citep{klimin2015}. For clarity of discussion, we postpone the results for nonzero temperature to Sec. \ref{ssec:finite-temperature}. 

We obtained the radial profiles $\rho_s(r)$ and $\rho_n(r)$ by substituting the numerical solution of Eq. (12) into Eqs. (6) and (7).
The numerical results in the BEC regime [$(k_Fa_s)^{-1}=1$], at unitarity [$(k_Fa_s)^{-1}=0$], and in the BCS regime [$(k_Fa_s)^{-1}=-1$] are shown in Fig. 1(a).
In all cases, the superfluid density varies smoothly between the center of the vortex ($r=0$) and the bulk ($r=\infty)$ density. On the BEC side of the crossover 
, the total density also vanishes at $r=0$. The normal density is small, but nonzero at finite $r$.
In contrast, on the BCS side, the total density is nonzero also in the center of the vortex. This implies the presence of a normal density at $r=0$, which becomes more pronounced when approaching the BCS limit. 
The asymptotics \eqref{eq:asymptotic-superfluid} and \eqref{eq:asymptotic-normal} are also plotted outside the core region, i.e. for $r > \xi$. In this region, they provide a good approximation for the densities.

\section{Vortex mass along the BEC--BCS crossover} \label{sec:vortex-mass}

Now that the densities are given, we can compute the vortex mass by integrating Eqs. \eqref{eq:associated-mass} and \eqref{eq:internal-mass} numerically.  However, to clarify how the vortex mass depends on the system size, we first consider a simplified analytic model.

\subsection{System-size dependence of vortex mass}
To evaluate the vortex mass analytically, taking into account both the asymptotic behavior and the contribution from the vortex core region,
 we can approximate $\rho_s$ by
\begin{equation} \label{eq:simplified-rhos}
    \rho_s \approx \begin{cases}\frac{1}{2} \rho_{s,\infty}\frac{r^2}{\xi^2}& r \leq \xi\\
\rho_{s,\infty} \left(1-\frac{\xi^2}{2r^2}\right) & r > \xi\end{cases}.
\end{equation}
The short-range part is chosen such that the function vanishes at $r =0$ and such that the profile is smoothly connected at $r = \xi$. If we also match the normal density at $r = \xi$, we can approximate
\begin{equation} \label{eq:simplified-rhon}
    \rho_n \approx\begin{cases} \rho_{n,\infty} + \delta\rho_{n,\infty} \left(1-\frac{1}{2} \frac{r^2}{\xi^2}\right)& r \leq \xi\\
\rho_{n,\infty} + \delta\rho_{n,\infty} \frac{\xi^2}{2r^2} & r > \xi\end{cases}
\end{equation}
As shown in Fig. \ref{fig:profiles-densities-T0}(a), there are some differences from the numerical results in the vortex core region, but these differences do not significantly affect the integrated values of the masses.
In this approximation, integral \eqref{eq:associated-mass} is easy to evaluate
\begin{align} \label{eq:approximate-associated-mass}
    M_a &\approx 2\pi \rho_{s,\infty} \int_{0}^\xi dr \, r \left(1-\frac{1}{2}\frac{r^2}{\xi^2}\right) + 2\pi \rho_{s,\infty} \int_{\xi}^R dr \, r \frac{1}{2}\frac{\xi^2}{r^2} \nonumber \\
    &= \pi \xi^2 \rho_{s,\infty}  \log\left(\frac{R}{\alpha_{\text{ana}}  \xi}\right) 
\end{align}
where we introduced $\alpha_{\text{ana}} = e^{-3/4} \approx 0.47 $.
An identical reasoning for integral \eqref{eq:internal-mass} leads to
\begin{equation} \label{eq:approximate-internal-mass}  M_i \approx \pi \xi^2 \,\delta\rho_{n,\infty}  \log\left(\frac{R}{\alpha_{\text{ana}} \xi}\right).
\end{equation}
In these formulas, it was necessary to introduce a long-range cutoff $R$, which can be interpreted as the system size \footnote{Realistically, a physical system in a box trap has a region of size $\sim \xi$ where the order parameter goes to zero. Thus, the cutoff needs to be slightly smaller than the radius of the physical system.}.
The logarithmic divergence is a result of the $r^{-2}$ term in the asymptotic form of the superfluid and normal densities for $r \to \infty$. Hence, the same logarithmic divergence will arise without the simplifying assumptions \eqref{eq:simplified-rhos} and \eqref{eq:simplified-rhon}.

These formulas are reminiscent of the vortex energy, which is proportional to $\log(R/r_c)$ where $r_c$ is the short-range cutoff, which is of the order of the vortex core size and is sensitive to microscopic details. In our calculation, the role of $r_c$ is played by $\alpha_{\text{ana}} \xi$, which depends on the
profiles in the vortex core region. In the next section, we will show how to replace this parameter by a more accurate value.

\subsection{Short-range cutoff}

In the previous section, the asymptotic behavior was taken into account exactly, but the density profiles close to the center of the vortex were approximated.
To evaluate the short-range cutoff more precisely, we calculate integrals \eqref{eq:associated-mass} and \eqref{eq:internal-mass} numerically and define parameters $\alpha_a$ and $\alpha_i$ by the following formulas.
\begin{align}
    M_a &= \pi \xi^2 \rho_{s,\infty} \log\left(\frac{R}{\alpha_a \xi}\right) \\
    M_i &=
    \pi \xi^2  \delta\rho_{n,\infty} \log\left(\frac{R}{\alpha_i \xi}\right).
\end{align}
The total vortex mass is then given by $M_{\text{tot}} = M_a + M_i$.
A priori, $\alpha_a$ and $\alpha_i$ still depend on the system size $R$. However, the dependence is very weak: we verified that the deviation from a constant is less than 1\% for $R/\xi > 3$. Thus, we can safely replace $\alpha_a$ and $\alpha_i$ by the values obtained for a sufficiently large system size. 
Naively, these parameters introduce new length scales, $\alpha_a \xi$ and $\alpha_i \xi$ that characterize the vortex core size. However, we will soon show that $\alpha_a$ and $\alpha_i$ are always of order unity, meaning they do not define an independent length scale.

\subsection{BEC--BCS crossover}

\begin{figure*}
    \centering
    \includegraphics[width=\textwidth]{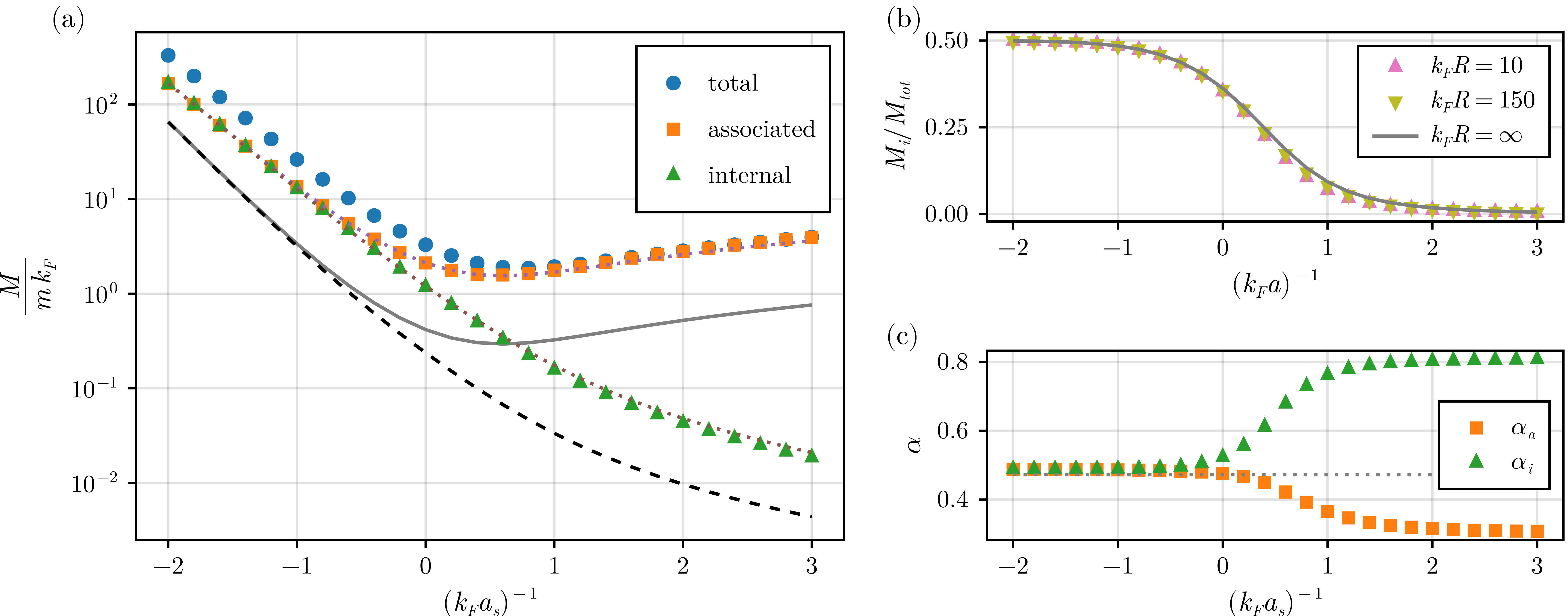}

    \caption{
    (a) The vortex mass for a system size of $k_FR = 150$ at zero temperature is plotted, alongside the contributions of the internal and associated mass.
    The dotted purple and brown lines show the simplified analytic model given by Eqs. \eqref{eq:approximate-associated-mass} and \eqref{eq:approximate-internal-mass}, respectively.  The solid gray line represents $\pi \xi^2 \rho_{s,\infty}$ and the dashed black line $\pi \xi^2 \,\delta\rho_{n,\infty}$.
    (b) The ratio of the internal vortex mass to the total vortex mass for different system sizes, including the $R\to\infty$ limit. (c) The correction factors $\alpha_a$ and $\alpha_i$. The dotted line shows the value from the analytical model $\alpha_{\text{ana}}$.}
    \label{fig:vortex_mass}
\end{figure*}

Let us now numerically evaluate integrals \eqref{eq:associated-mass} and \eqref{eq:internal-mass} along the crossover.
In the experiments \citep{kwon2021, delpace2022, hernandez-rajkov2024}, a system with radius $45 \,\mathrm{\mu m}$ is used and the typical Fermi wave vector is $k_F^{-1} \sim 0.3\,\mathrm{\mu m}$. Therefore, we choose to plot this for $k_F R = 150$.
The results are shown in Fig. \ref{fig:vortex_mass}. We see that the total mass sharply increases on the BCS side. It reaches a minimum around $(k_F a_s)^{-1} \sim 0.5$ and increases again toward the BEC limit.
The mass we obtained is significantly larger than $\pi \xi^2 \rho_{s,\infty}$, which is the local associated mass. For $k_F R = 150$, the total computed mass is about a factor 5 higher than this value, depending on the scattering length. This confirms that the logarithmic part dominates for realistic system sizes.
We can also look at the prefactor of the logarithm of the internal mass $\pi \xi^2 \,\delta\rho_{n,\infty}$. This quantity decreases monotonically from the BCS side to the BEC side. 
We note that in our simplified analytical model
Eqs. \eqref{eq:simplified-rhos}, \eqref{eq:simplified-rhon}, the normal density in the center is $\rho_n(r=0) = \rho_{n,\infty} + \delta\rho_{n,\infty}$. In this approximation, the prefactor $\pi \xi^2 \,\delta\rho_{n,\infty}$ is the normal component that exceeds the background value in the center of the vortex and can hence be considered a local internal mass. 

We can examine the vortex mass as a function of the (inverse) scattering length more closely by looking at contributions of the associated and internal masses separately. While the internal mass becomes very small in the BEC limit, the associated mass increases. This is caused by the healing length increasing toward the BEC limit. In the BCS regime, it can be seen that the associated and internal masses coincide. This is at first sight quite surprising. However, we can analyze this from the density profiles given in Fig. \ref{fig:profiles-densities-T0}. It can be seen that the total density approaches a constant value in the BCS limit. In this limit, the values of integrals \eqref{eq:associated-mass} and \eqref{eq:internal-mass} are the same.
This can also be seen in the correction factors, which are plotted in Fig. \ref{fig:vortex_mass}. In the BCS both $\alpha_a$ and $\alpha_i$ go to the same value, which is in fact very close to our analytic prediction. The results of the analytic model are plotted with dotted lines in Fig. \ref{fig:vortex_mass}(a). This agrees quantitatively with the numerical result in the BCS limit, but slightly underestimates the associated mass and overestimates the internal mass in the BEC limit. This can be solved by using the values of $\alpha_a$ and $\alpha_i$ given in Fig. \ref{fig:vortex_mass}(c) instead.
Furthermore, it is interesting to note how the two parts contribute to the total vortex mass as a function of the inverse scattering length. In the $R\to\infty$ limit, the ratio of the internal to the total mass can be calculated as
\begin{align}
\lim_{R\to \infty} \frac{M_i}{M_{\text{tot}}} =
\frac{\rho_{s,\infty}}{2\rho_{s,\infty} - \Delta^2   \left.\pdv{\rho}{ |\Phi|^2}\right|_\infty}
\end{align}
For this quantity, the logarithmic divergences cancel.
As can be seen from Fig. \ref{fig:vortex_mass}, this $R\to\infty$ result already works well for modest system sizes.
Again, we reach a similar conclusion: the internal mass goes to zero on the BEC side but becomes equal to the associated mass on the BCS side. This can be seen more explicitly by noting that $\partial \rho/ \partial|\Phi|^2|_\infty$ vanishes in the BCS limit.

\subsection{Results for nonzero temperature}
\label{ssec:finite-temperature}

\begin{figure}
    \centering
\includegraphics[width=\columnwidth]{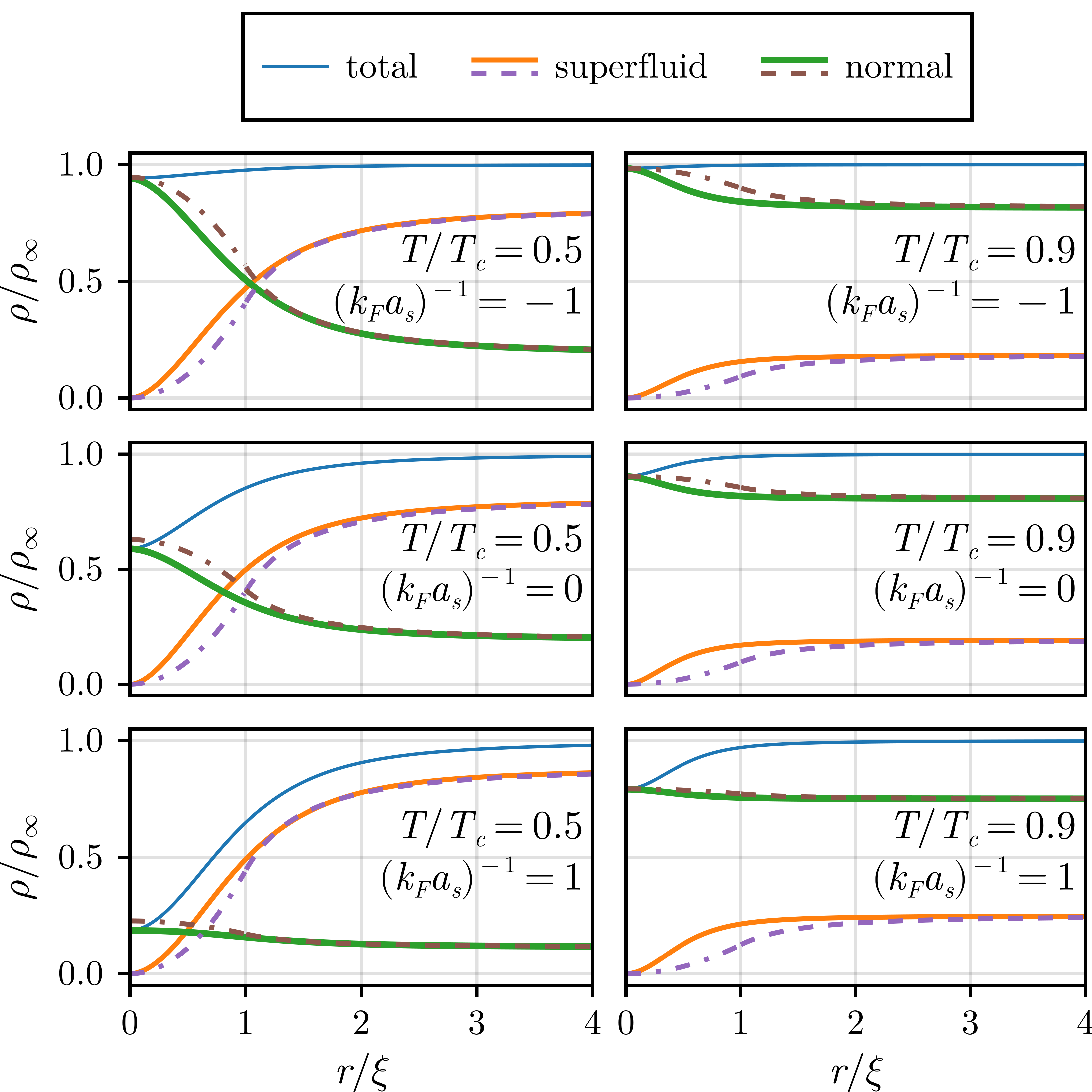}
    \caption{
    Total density $\rho_{\text{tot}}$ (solid blue line), superfluid density $\rho_s$ (solid orange line) and normal density $\rho_n$ (solid green line) as a function of the radial coordinate, calculated in the EFT framework at nonzero temperatures. The densities are normalized by the total density at $r=\infty$.
    Also plotted are the asymptotic expressions for the superfluid (dashed purple line) and normal (dashed brown line) densities. The gray dotted lines correspond to the simplified analytic vortex profile introduced in Eqs. \eqref{eq:simplified-rhos} and \eqref{eq:simplified-rhon}.}
    \label{fig:densities-finT}
\end{figure}

\begin{figure*}
    \centering
    \includegraphics[width=\textwidth]{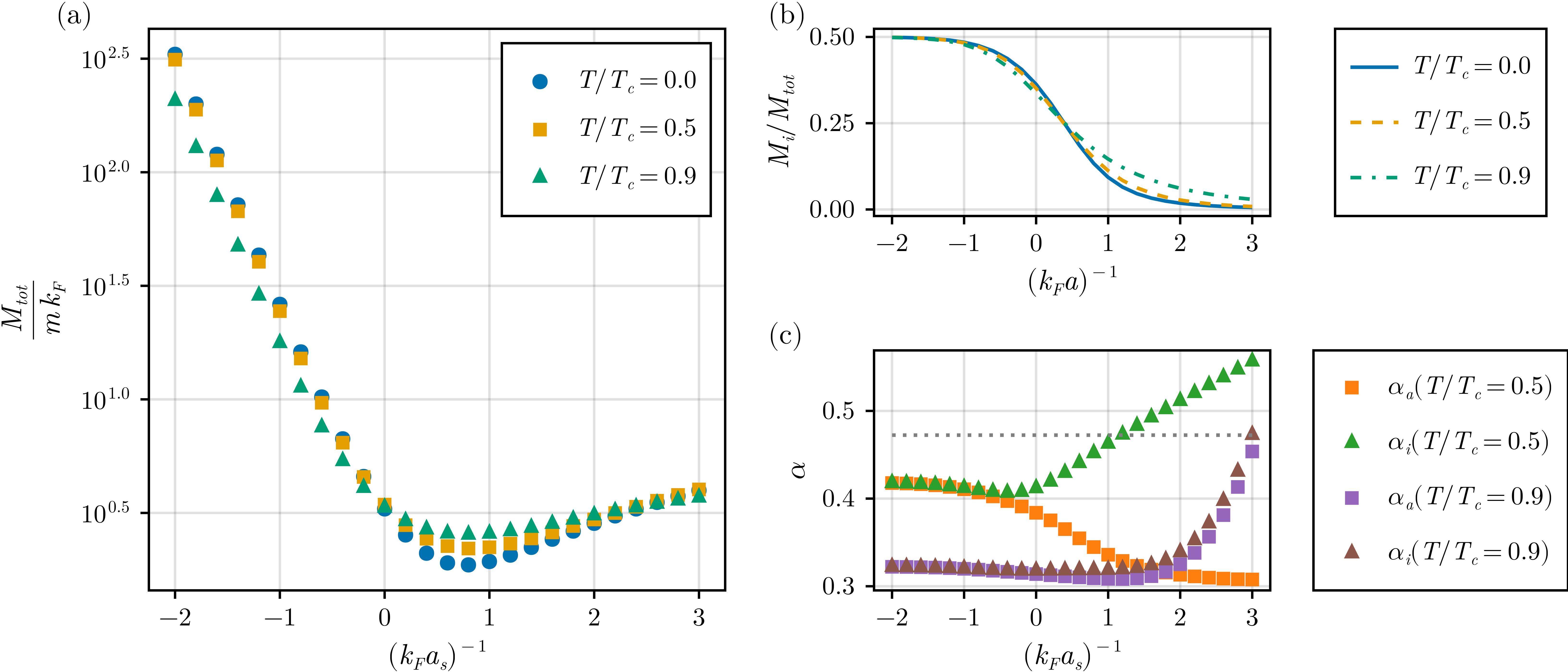}
  
    \caption{ (a) The (total) vortex mass for a system size of $k_FR = 150$ at various temperatures.
    (b) The ratio of the internal vortex mass to the total vortex mass at various temperatures in the $R\to\infty$ limit.
    (c)  The correction factors $\alpha_a$ and $\alpha_i$ at temperatures $T/T_c = 0.5$ and $T/T_c = 0.9$.}
    \label{fig:vortex_mass-finT}
\end{figure*}

If the temperature is nonzero, a normal density at infinity appears for all interaction strengths, as can be seen in Fig. \ref{fig:densities-finT}. Close to the critical temperature, most of the density is in fact located in this normal density, also in the BEC case.
Then, it is important to subtract $\rho_{n,\infty}$ in Eq. \eqref{eq:internal-mass}. This makes physical sense, because the origin of the internal mass is the fact that normal component needs to be moved, hence not including the normal density that is equally distributed over the entire system.
The results for a system size of $k_F R = 150$ are plotted in Fig. \ref{fig:vortex_mass-finT}.
We can see that the impact of the temperature depends on the interaction strength.
On the BCS side, the vortex mass decreases with increasing temperature. In contrast, around unitarity and on the BEC side, the mass increases. 

We can explain this quite simply by noting that the associated mass is proportional to $\pi \xi^2 \rho_{s,\infty}$. As the temperature is raised, the healing length increases but the superfluid density decreases. On the BCS side, the lowering of the superfluid density leads to a decrease. On the BEC side, the increase of the healing length instead leads to an increase. A similar argument holds for the internal mass, replacing $\rho_{s,\infty}$ by $\delta\rho_{n,\infty}$. On the BCS side, also the internal mass decreases, which can be understood as the increasing background normal state density resulting in a decrease of
the \emph{excess} normal component density in the vortex core.

It is also interesting to look at the correction factors $\alpha_a$ and $\alpha_i$ at various temperatures. For temperature 0.5$T_c$, a similar pattern as in the zero-temperature case emerges. In the BCS limit both correction factors become equal, while they are different on the BEC side. In contrast, at temperature 0.9$T_c$ the correction factors stay similar even on the BEC side. This should be interpreted as a result of the vortex core structure being significantly different from the zero temperature case.

\section{Conclusion and discussion}

In this paper, we studied the vortex mass problem in an effective field theory framework.
We derived that the associated and internal masses are logarithmically dependent on the system size. In contrast, these were considered finite and of order $\pi \rho_{s,\infty} \xi^2$ in previous literature \citep{baym1983, kopnin1978, simula2018, richaud2024}.
This prefactor still sets the general scale, but needs to be multiplied by a factor of about 5 for a realistic system size of $ k_F R = 150$.
We note that, since the healing length in the BEC regime is proportional to $(k_F a_s)^{-1}$, the vortex mass in fact also increases in the BEC limit, even though the internal mass goes to zero. 
In the case of a BEC, it is known that massless point-vortex models provide a good approximation to the dynamics \citep{navarro2013, samson2016}. Nevertheless, precision experiments in this regime may be able to observe the vortex mass.
We should note that this conclusion is quite general: the logarithmic dependence will be valid as long as the order parameter vortex profile has the asymptotic behavior $f \sim 1 - \xi^2/4r^2$, which is also predicted by Gross--Pitaevskii and Bogoliubov--de Gennes theories. The scale of the vortex mass is set by the prefactor of the logarithm. For the associated mass, this prefactor is simply $\pi \rho_{s,\infty} \xi^2$. This quantity may also be evaluated in other theoretical frameworks, $\xi$ being defined by the relation above. The other quantities appearing in the vortex mass are the correction factors $\alpha_a$ and $\alpha_i$. As we derived, these factors are always of order unity. These factors may be affected by details of the core structure; however, they do not have a large impact on the order of magnitude of the vortex mass. Nevertheless, quantitative improvements may be achieved using a more microscopic approach of calculating the normal state core density, especially including CdGM states.
We showed that the internal mass goes to zero in the BEC limit and becomes equal to the associated mass in the BCS limit. Also this conclusion is independent of the details of the model. Along the crossover, the value interpolates between these two limits.  Conceptually, the independence of the vortex structure is in line with the assumptions underlying the point-vortex model, which does not consider the details of the vortex core.
Another quantitative improvement could be made by adding beyond-mean-field corrections. Changing the equation of state allows to solve the problem that the dimer--dimer scattering length in the BEC limit is wrongly predicted to be $2a_s$, whereas more careful analysis indicates $0.6a_s$, as has been discussed in the literature \citep{petrov2004, hu2006}. However, we expect that the qualitative features of the results will not change, and that our mean-field results can already provide a useful benchmark for future experiments and beyond-mean-field theories.

We may compare our results to those previously obtained in the literature. First, note that we disregarded the relativistic or compressibility mass in our treatment. However, it is interesting to note that the expressions in the literature coincide with our result, apart from the correction factor. 
This opens up the question whether these should be seen as distinct, or whether they can be considered the same.
Next, let us compare to the recent results by \citet{richaud2024}. Their derivation contains some parallels to ours. They also conclude that most of the internal mass is \emph{not} caused by the CdGM states, as was assumed by Kopnin.
In contrast to our derivation, they do not take into account the associated mass but only the internal mass. Furthermore, they do not consider the system-size dependence. That being said, the internal mass is computed in the same way (also considering the background subtraction at nonzero temperatures), but starting from BdG solutions.
This calculation of the vortex mass is compared to a numerical simulation of the dynamics of a single vortex using a time-dependent density functional theory method. The vortex mass is extracted from the simulations by looking at the oscillation frequency of the vortex. In contrast to our calculations, the derivation is only done for a limited parameter regime in the BCS limit. Still, it is quite striking that they find good agreement with their time-evolution scheme, even when disregarding the associated mass.  In contrast, our approach would suggest a mass of about a factor of 2 higher. This would lead to an amplification of the transverse oscillations of the vortex compared to the ones observed in the numerical simulations in Ref. \citep{richaud2024}. We note that on BCS side our model predicts that the vortex mass decreases, whereas in Ref. \citep{richaud2024} an increase was predicted.
Some further analysis is needed to clear up these discrepancies. 
Our results may be put in the more general context of multicomponent superfluids; indeed we used a two-fluid description, in which the normal component plays the role of the internal mass of the vortex. A similar picture is expected to arise for mixtures of Bose gases. There the minority component will accumulate in the vortex core leading to a contribution to the internal mass.

Our results have important implications for the measurement of the vortex mass.
We propose to measure the vortex mass in an experimental setup similar to Refs. \citep{kwon2021, delpace2022, hernandez-rajkov2024}. The vortex mass can be accurately determined by creating a vortex pair and looking at the rotation frequency, as discussed in Ref. \citep{kanjo2024}. Because of the system-size dependence, our prediction is significantly higher than other proposals. Moreover, the logarithmic dependence could be verified explicitly by varying the system size. We should take some care when extrapolating our results for a single vortex to the multi-vortex case. In particular, the vortices should be located close enough to the center. The correction to the vortex mass from displacing the vortex by $\delta r$ is of order  $(\delta r)^2/ R^2$ times the prefactor of the logarithm, which should be a small correction for large-size systems.


In conclusion, we found a logarithmic dependence of the associated and internal masses on the system size, and characterized it by the healing length $\xi$ and correction factors $\alpha_a$, $\alpha_i$.
We evaluated the vortex mass over the BEC--BCS crossover, both at zero and at nonzero temperature. We find that the vortex mass increases sharply on the BCS side, but also more slowly in the BEC limit. Because of the logarithmic dependence, the vortex mass is significantly larger than expected from a naive estimate. This has important consequences for measurability of the vortex mass in state-of-the-art experiments.

\begin{acknowledgments}
We thank Serghei Klimin for fruitful discussions. L.L. and J.T. acknowledge financial support by the Research Foundation Flanders (FWO), Projects No. G0AIY25N, No. G0A9F25N and No. G060820N.
H.T. is supported by JSPS KAKENHI Grants No. JP18KK0391, and No. JP20H01842; and JST, PRESTO (Japan) Grant No. JPMJPR23O5.
\end{acknowledgments}

\section*{Data Availability}
The data that support the findings of this article are not
publicly available. The data are available from the authors
upon reasonable request.

\appendix


\section{Massive point-vortex model} \label{app:pvm}

In a thin disk geometry, vortices are constrained to have their axes aligned perpendicular to the plane of the disk. The vortices can be considered as small filaments, and the degrees of freedom that do not correspond to in-plane motion can be neglected. In this context, a common model to describe the dynamics of vortices is the point-vortex model, in which the vortices are approximated by point particles exerting forces on each other. Each vortex has a position $\mathbf r_j$ and a vortex charge $\Gamma_j$, equal to an integer multiple of the circulation quantum $h/(2m)$, leading to the following velocity field
\begin{equation}
    \mathbf v_j(\mathbf r) =  \frac{\Gamma_j}{2 \pi} \frac{\mathbf{r}_j-\mathbf{r}}{\left|\mathbf{r}_j-\mathbf{r}\right|^2}
\end{equation}
around each vortex. Each vortex is assumed to move in the velocity field caused by all other vortices, undergoing a Magnus force.
The Magnus force (per unit length of the vortex filament) on the $i$the vortex is given by \citep{saffman1993}
\begin{equation}
    \mathbf f_i =\rho_s \Gamma_i \hat{\mathbf{e}}_z \times\left(\frac{d \mathbf{r}_i}{d t}-\hat{\mathbf{e}}_z \times \sum_{j \neq i} \frac{\Gamma_j}{2 \pi} \frac{\mathbf{r}_i-\mathbf{r}_j}{\left|\mathbf{r}_i-\mathbf{r}_j\right|^2}\right).
\end{equation}
When the vortices are assumed massless, the equation of motion is $\mathbf f_i = 0$, more succinctly written as
\begin{equation}
\frac{d \mathbf{r}_i}{d t}=\hat{\mathbf{e}}_z \times \sum_{j \neq i} \frac{\Gamma_j}{2 \pi} \frac{\mathbf{r}_i-\mathbf{r}_j}{\left|\mathbf{r}_i-\mathbf{r}_j\right|^2} \label{eq:massless-pvm}
\end{equation}
which is often used in the literature. If the vortex mass is nonzero, we have the equation of motion
\begin{equation}
M \frac{d^2 \mathbf{r}_i}{d t^2}=\rho_s \Gamma_i \hat{\mathbf{e}}_z \times\left(\frac{d \mathbf{r}_i}{d t}-\hat{\mathbf{e}}_z \times \sum_{j \neq i} \frac{\Gamma_j}{2 \pi} \frac{\mathbf{r}_i-\mathbf{r}_j}{\left|\mathbf{r}_i-\mathbf{r}_j\right|^2}\right),
\end{equation}
where $M$ is the vortex mass per unit length of the vortex filament. In our model, we consider the associated and internal masses as contributing independently to the vortex mass, so $M = M_a + M_i$.
When $M$ is set to zero, we recover the trajectories obtained from the massless point-vortex model \eqref{eq:massless-pvm}. Including the mass term leads to oscillations on top of these trajectories, with a frequency proportional to the cyclotron frequency
$\Omega_c = \frac{\rho_s (h/2m)}{M}$ and amplitude proportional to the vortex mass length scale $\sigma = \sqrt{\frac{M}{\pi \rho_s}}$, as introduced in \citep{kanjo2024}. As the vortex mass increases, the amplitude of these oscillations increases and their frequency decreases. In the (singular) limit $M \to 0$ the frequency diverges but the amplitude of the oscillations goes to zero.

There is an electromagnetic analogy that can be made explicit introducing the effective Hamiltonian $\mathcal{H}_\text{vort} = \sum_j \tfrac{(\mathbf{p}_j-q\mathbf{A}(\mathbf r_j))^2}{2M} + H_0$ with the canonical momentum $\mathbf{p}_j=M\dot{\mathbf{r}}_j+q\mathbf{A}(\mathbf r_j)$ and the position $\mathbf{r}_j$ of the $j$th vortex.
Here, $H_0$ is the Hamiltonian (energy) of the point-vortex model without vortex inertia, and $q_j=\sqrt{\epsilon_0 \rho_s} \Gamma_j$ the effective charge and $\mathbf{A}(\mathbf r)=\sqrt{\rho_s/\epsilon_0} \mathbf{r}\times \mathbf{e}_z$ the effective vector potential with the dielectric constant $\epsilon_0$ in vacuum based on the correspondence with the two-dimensional Coulomb gas (see for example \citep{kanjo2024}).

\section{Effective field theory} \label{app:eft}

We briefly sketch the derivation of the effective field theory and provide explicit expressions for the coefficients appearing in the action functional and the equation of motion.

The starting point of the EFT is the Euclidean action of a Fermi gas interacting via an $s$-wave contact interaction
\begin{align*}
    S_E[\psi, \psi^*] &= \int d\tau d\mathbf{x} \Big[ \sum_{\sigma = \uparrow,\downarrow} \psi^*_{\sigma}\left(\partial_\tau - \frac{1}{2m}\nabla^2 - \mu_\sigma\right) \psi_\sigma \\ 
    &\hspace{5em} + g \psi_\downarrow^* \psi_\uparrow^* \psi_\uparrow \psi_\downarrow \Big]
\end{align*}
For equal spin populations $\mu_\uparrow = \mu_\downarrow$. More generally we can define an average chemical potential $\mu = (\mu_\uparrow + \mu_\downarrow)/2$ and an imbalance chemical potential $\zeta = (\mu_\uparrow - \mu_\downarrow)/2$.
Now the BCS order parameter $\Phi$ is introduced using the Hubbard--Stratonovich transformation. After integrating out the fermions, the result is
\begin{equation*}
    S[\Phi, \bar \Phi] = -\int_0^{\hbar\beta} d\tau \int d\mathbf x \frac{\bar \Phi(\tau, \mathbf x) \Phi(\tau, \mathbf x)}{g}-\operatorname{Tr}\left[\ln \left(-\mathbb{G}^{-1}\right)\right],
\end{equation*}
where
\begin{align*}
\bra{\tau', \mathbf x'} - \mathbb G^{-1} \ket{\tau, \mathbf x} &= \delta(\tau-\tau') \delta(\mathbf x-\mathbf x') \\
&\quad\begin{pmatrix}
\partial_\tau-\nabla^2-\mu-\zeta & -\Phi(\boldsymbol{r}, \tau) \\
-\Phi^*(\boldsymbol{r}, \tau) & \partial_\tau+\nabla^2+\mu-\zeta
\end{pmatrix},
\end{align*}
and $\ln (\mathbb A) = \sum_{n=0}^\infty \mathbb (1 - \mathbb A)^n /n $ is to be regarded as a formal power series.
This transformation is exact, but the trace appearing in this action cannot be carried out explicitly. However, the trace can be evaluated using a gradient expansion, which is valid if the pair correlation length is small with respect to the healing length \citep{lombardi2016}. The details of this calculation are nontrivial, and are presented in Refs. \cite{klimin2015, lombardi2017}. 
The result is the action
\begin{multline*}
    S_{EFT}[\Phi,\bar\Phi] =\\
    \int_0^{\hbar \beta} d\tau \int d \mathbf{x} \left[\hbar\frac{D(|\Phi|^2)}{2}\left(\bar \Phi\frac{\partial\Phi}{\partial\tau}- \frac{\partial\bar \Phi}{\partial\tau}\Phi\right) + \hbar^2 Q \frac{\partial \bar \Phi}{\partial \tau}  \frac{\partial \Phi}{\partial \tau} \right.\\
    - \hbar^2 R\left(\frac{\partial{|\Phi|^2}}{\partial \tau}\right)^2 
+ \frac{\hbar^2C}{2m} \left(\nabla \bar \Phi\cdot \nabla \Phi\right)\\ \left.
- \frac{\hbar^2 E}{2m} (\nabla |\Phi|^2)^2
+ \Omega_{sp}(|\Phi|^2)
\right].
\end{multline*}
We now give the explicit expressions for the coefficients appearing in this action.
As before, define
\begin{align*}
\xi_{\mathbf k} = \frac{\hbar^2 k^2}{2m} - \mu \\
E_{\mathbf k}(|\Phi|^2) = \sqrt{\xi_{\mathbf k}^2 + |\Phi|^2}.
\end{align*}
For compactness, we also write $E_{\mathbf k} = E_{\mathbf k}(\Delta)$ where $\Delta$ is the value of the order parameter in the bulk.
We further define the functions $f_n$ recursively by

\begingroup
\allowdisplaybreaks
\begin{align*}
f_1(\beta,\epsilon,\zeta) &\coloneqq \frac{X(\epsilon)}{2\epsilon} = \frac{1}{2\epsilon} \frac{\sinh(\beta\epsilon)}{\cosh(\beta\epsilon)+\cosh(\beta\zeta)} \\
f_{n+1}(\beta, \epsilon, \zeta) &\coloneqq -\frac{1}{2n\epsilon} \frac{\partial f_n}{\partial \epsilon}.
\end{align*}
The EFT coefficients are given by
\begin{align*}
&\Omega_{sp}(|\Phi|^2)=-\frac{m}{4 \pi  \hbar^2 a_s}|\Phi|^2-\int \frac{d \mathbf{k}}{(2 \pi)^3}\Big[-\xi_{\mathbf{k}}-\frac{m}{\hbar^2k^2} |\Phi|^2\nonumber\\
&\hspace{7em}+ \frac{1}{\beta} \ln (2 \cosh \left(\beta E_{\mathbf{k}}(|\Phi|^2)\right)+2\cosh (\beta \zeta)) \Big] \\
& \mathcal A(|\Phi|^2)=-\frac{m}{4\pi\hbar^2 a_s}\\
&\hspace{7em}-\int \frac{d \mathbf{k}}{(2 \pi)^3} \left( f_1\left(\beta, E_{\mathbf{k}}(|\Phi|^2), \zeta\right) - \frac{m}{\hbar^2 k^2}\right) \\
&D(|\Phi|^2)=\int \frac{d \mathbf{k}}{(2 \pi)^3} \frac{\xi_{\mathbf{k}}}{|\Phi|^2}\left[f_1\left(\beta, \xi_{\mathbf{k}}, \zeta\right)-f_1\left(\beta, E_{\mathbf{k}}(|\Phi|^2), \zeta\right)\right] \\
& \tilde D(|\Phi|^2)=\int \frac{d \mathbf{k}}{(2 \pi)^3} \xi_{\mathbf{k}} f_2\left(\beta, E_{\mathbf k}(|\Phi|^2), \zeta\right) \\
& C=\int \frac{d \mathbf{k}}{(2 \pi)^3} \frac{\hbar^2 k^2}{3 m} f_2\left(\beta, E_{\mathbf{k}}, \zeta\right) \\
& E=2 \int \frac{d \mathbf{k}}{(2 \pi)^3} \frac{\hbar^2k^2}{3 m} \xi_{\mathbf{k}}^2 f_4\left(\beta, E_{\mathbf{k}}, \zeta\right) \\
& Q=\frac{1}{2\Delta^2} \int \frac{d \mathbf{k}}{(2 \pi)^3}\left[f_1\left(\beta, E_{\mathbf{k}}, \zeta\right)-\left(E_{\mathbf{k}}^2+\xi_{\mathbf{k}}^2\right) f_2\left(\beta, E_{\mathbf{k}}, \zeta\right)\right] \\
& R= \int \frac{d \mathbf{k}}{(2 \pi)^3}\Bigg[\frac{f_1\left(\beta, E_{\mathbf{k}}, \zeta\right)+\left(E_{\mathbf{k}}^2-3 \xi_{\mathbf{k}}^2\right) f_2\left(\beta, E_{\mathbf{k}}, \zeta\right)}{6\Delta^4} \nonumber \\
&\quad+\frac{2\left(\xi_{\mathbf{k}}^2-2 E_{\mathbf{k}}^2\right)}{3\Delta^2} f_3\left(\beta, E_{\mathbf{k}}, \zeta\right)+ E_{\mathbf{k}}^2 f_4\left(\beta, E_{\mathbf{k}}, \zeta\right)\Bigg]. \\
G &= 
\int \frac{d \mathbf{k}}{(2 \pi)^3} f_2\left(\beta, E_{\mathbf{k}}, \zeta\right)
\end{align*}
\endgroup

The quantities $\rho_{s,\infty}$ and $\delta\rho_{n,\infty}$ can be expressed in terms of these integrals
\begin{align}
\rho_{s,\infty} &= 4m C \Delta^2 \\
\delta\rho_{n,\infty} &=  \rho_{s,\infty}-2m\Delta^2 \tilde D(\Delta^2).
\end{align}

\section{BCS and BEC limits of the healing length} \label{app:healing}

In \citet{marini1998}, various quantities were evaluated analytically at zero temperature using elliptic integrals. We can follow an analog approach for the healing length
\begin{equation}
    \xi = \sqrt{\frac{\hbar^2}{m}\frac{C}{\Delta^2G}}.
\end{equation}
At zero temperature, the EFT coefficients $C$ and $G$
 take the form
\begin{align}
    C=\int \frac{d \mathbf{k}}{(2 \pi)^3} \frac{\hbar^2 k^2}{3 m}  \frac{1}{2E_{\mathbf{k}}^3}
     &= \frac{1}{2\pi^2} \frac{
     (2m)^{3/2}}{\hbar^3\Delta^{1/2}} \frac{I_2(\mu/\Delta)}{4}
    \\
    G=\int \frac{d \mathbf{k}}{(2 \pi)^3} \frac{1}{2E_{\mathbf{k}}^3}
    &= \frac{1}{2\pi^2} \frac{
     (2m)^{3/2}}{\hbar^3 \Delta^{3/2}} \frac{I_5(\mu/\Delta)}{8},
\end{align}
where $I_2$ and $I_5$ are the integrals calculated in Ref. \citep{marini1998}. Using the asymptotic forms
\begin{align}
 I_2(x) &\sim \begin{cases}
         \frac{2}{3} x^{3/2}& x \to \infty \\
         \frac{\pi}{8} (-x)^{-1/2}& x \to -\infty
    \end{cases} \\
I_5(x) &\sim \begin{cases}
         x^{1/2}& x \to \infty \\
         \frac{\pi}{16} (-x)^{-3/2}& x \to -\infty
    \end{cases}
\end{align}
and
\begin{align}
    \frac{\mu}{E_F} \sim \begin{cases}
        1 & \text{(BCS limit)} \\
        -(k_F a_s)^{-2} & \text{(BEC limit)}
    \end{cases},
\end{align}
we obtain
\begin{equation}
    \xi \sim \begin{cases}
        \sqrt{\frac{\hbar^2}{m\Delta} \frac{2x}{3}} =
        \frac{1}{\sqrt{3}} \frac{\hbar^2 k_F}{m\Delta}
        & \text{(BCS limit)} \\
         \sqrt{\frac{\hbar^2}{m\Delta} (-2x)} =
         \sqrt{2} (k_F a_s)\frac{\hbar^2 k_F}{m\Delta}
         & \text{(BEC limit)}
    \end{cases}
\end{equation}
In the BEC case, this can be further rewritten as
\begin{equation}
    \xi \sim \frac{\hbar}{\sqrt{\rho g_B}},\quad g_B = \frac{4\pi \hbar^2 \cdot 2a_s}{2m},
\end{equation}
which is the standard definition of the bosonic healing length for a gas of bosons of mass $2m$ and $s$-wave scattering length $2a_s$, consistent with expectations from mean-field theory. Note that many references use an alternative definition rescaling by a factor $\sqrt{2}$. \cite{pethick2006, pitaevskii2003}

Up to a prefactor, the results in the BCS limit also agree with mean-field results in the literature \citep{marini1998}.

\section{Details of numerical algorithm for the vortex profile} \label{app:num}
To solve Eq. \eqref{eq:eft-profile-equation} we use an imaginary time evolution method
\begin{multline}
- \pdv{f}{\tau'} = -\frac{\hbar^2C}{2m} \Delta \left(\frac{1}{r}\pdv{}{r}\left[r \pdv{f}{r}\right] - \frac{q^2}{r^2} f(\tau', r)\right) \\
+ \Delta^3\frac{\hbar^2E}{m} \frac{f(\tau', r)}{r} \pdv{}{r}\left[r \pdv{}{r} f^2(\tau', r)\right]\\
+ \mathcal A( \Delta^2 f^2(\tau', r))\Delta f(\tau', r).
\end{multline}
Note that unlike for the Gross--Pitaevskii case, this imaginary time coordinate $\tau'$ does not correspond to the time coordinate in the full equation of motion, and is only introduced as a numerical trick. We use a Dirichlet boundary condition on the left side of the domain and a Neumann boundary condition on the right sides. The derivatives are replaced by central differences on a discretized line with spacing $\xi/8$. We use a box size of $R = 150\xi$.
Starting from $f(\tau' = 0, r) = \tanh(r/\xi_1)$, we evolve until the solution converges to the steady state.


\bibliography{references_final.bib}

@article{baym1983,
  title = {The Hydrodynamics of Rotating Superfluids. {{I}}. {{Zero-temperature}}, Nondissipative Theory},
  author = {Baym, Gordon and Chandler, Elaine},
  year = 1983,
  month = jan,
  journal = {Journal of Low Temperature Physics},
  volume = {50},
  number = {1},
  pages = {57--87},
  issn = {1573-7357},
  doi = {10.1007/BF00681839},
  urldate = {2024-11-14},
  langid = {english}
}

@article{bellettini2023,
  title = {Relative Dynamics of Quantum Vortices and Massive Cores in Binary {{BECs}}},
  author = {Bellettini, Alice and Richaud, Andrea and Penna, Vittorio},
  year = 2023,
  month = aug,
  journal = {The European Physical Journal Plus},
  volume = {138},
  number = {8},
  pages = {676},
  issn = {2190-5444},
  doi = {10.1140/epjp/s13360-023-04294-6},
  urldate = {2024-09-10},
  langid = {english}
}

@article{berthod2016,
  title = {Vortex Spectroscopy in the Vortex Glass: {{A}} Real-Space Numerical Approach},
  shorttitle = {Vortex Spectroscopy in the Vortex Glass},
  author = {Berthod, C.},
  year = 2016,
  month = nov,
  journal = {Physical Review B},
  volume = {94},
  number = {18},
  pages = {184510},
  issn = {2469-9950, 2469-9969},
  doi = {10.1103/PhysRevB.94.184510},
  urldate = {2025-03-03},
  copyright = {http://link.aps.org/licenses/aps-default-license},
  langid = {english}
}

@article{botelho2006,
  title = {Vortex-{{Antivortex Lattice}} in {{Ultracold Fermionic Gases}}},
  author = {Botelho, S. S. and {S{\'a} de Melo}, C. A. R.},
  year = 2006,
  month = feb,
  journal = {Physical Review Letters},
  volume = {96},
  number = {4},
  pages = {040404},
  issn = {0031-9007, 1079-7114},
  doi = {10.1103/PhysRevLett.96.040404},
  urldate = {2024-11-05},
  copyright = {http://link.aps.org/licenses/aps-default-license},
  langid = {english}
}

@article{chen1994,
  title = {Shooting Method for Vortex Solutions of a Complex-Valued {{Ginzburg}}--{{Landau}} Equation},
  author = {Chen, Xinfu and Elliott, Charles M. and Qi, Tang},
  year = 1994,
  journal = {Proceedings of the Royal Society of Edinburgh: Section A Mathematics},
  volume = {124},
  number = {6},
  pages = {1075--1088},
  issn = {0308-2105, 1473-7124},
  doi = {10.1017/S0308210500030122},
  urldate = {2025-03-05},
  copyright = {https://www.cambridge.org/core/terms},
  langid = {english}
}

@article{delpace2022,
  title = {Imprinting {{Persistent Currents}} in {{Tunable Fermionic Rings}}},
  author = {Del Pace, G. and Xhani, K. and Muzi Falconi, A. and Fedrizzi, M. and Grani, N. and Hernandez Rajkov, D. and Inguscio, M. and Scazza, F. and Kwon, W. J. and Roati, G.},
  year = 2022,
  month = dec,
  journal = {Physical Review X},
  volume = {12},
  number = {4},
  pages = {041037},
  publisher = {American Physical Society},
  doi = {10.1103/PhysRevX.12.041037},
  urldate = {2025-03-20}
}

@book{donnelly1991,
  title = {Quantized Vortices in Helium {{II}}},
  author = {Donnelly, Russell J.},
  year = 1991,
  series = {Cambridge Studies in Low Temperature Physics},
  edition = {1. publ},
  number = {3},
  publisher = {Cambridge University Press},
  address = {Cambridge},
  isbn = {978-0-521-32400-7},
  langid = {english}
}

@article{duan1992,
  title = {Inertial {{Mass}} of a {{Moving Singularity}} in a {{Fermi Superfluid}}},
  author = {Duan, Ji-Min and Legett, Anthony J},
  year = 1992,
  journal = {Physical Review Letters},
  volume = {68},
  number = {8},
  pages = {1216--1219},
  doi = {10.1103/PhysRevLett.68.1216}
}

@article{duan1994,
  title = {Mass of a vortex line in superfluid  $^{4}\mathrm{He}$: Effects of gauge-symmetry breaking},
  shorttitle = {Mass of a vortex line in superfluid <span class="aps-inline-formula"><math xmlns="http},
  author = {Duan, Ji-Min},
  year = 1994,
  journal = {Physical Review B},
  volume = {49},
  number = {17},
  pages = {12381-12383},
  doi = {10.1103/PhysRevB.49.12381}
}

@article{fil2007,
  title = {Mass of an {{Abrikosov}} Vortex},
  author = {Fil, V. D. and Ignatova, T. V. and Burma, N. G. and Petrishin, A. I. and Fil, D. V. and Shitsevalova, N. {\relax Yu}.},
  year = 2007,
  month = dec,
  journal = {Low Temperature Physics},
  volume = {33},
  number = {12},
  pages = {1019--1022},
  issn = {1063-777X},
  doi = {10.1063/1.2747080},
  urldate = {2025-02-27}
}

@article{franzen2022,
  title = {Observation of magnetic Feshbach resonances between Cs and $^{173}\mathrm{Yb}$},
  author = {Franzen, Tobias and Guttridge, Alexander and Wilson, Kali E. and Segal, Jack and Frye, Matthew D. and Hutson, Jeremy M. and Cornish, Simon L.},
  year = 2022,
  month = oct,
  journal = {Physical Review Research},
  volume = {4},
  number = {4},
  pages = {043072},
  publisher = {American Physical Society},
  doi = {10.1103/PhysRevResearch.4.043072},
  urldate = {2025-04-15}
}

@article{giorgini2008,
  title = {Theory of Ultracold Atomic {{Fermi}} Gases},
  author = {Giorgini, Stefano and Pitaevskii, Lev P. and Stringari, Sandro},
  year = 2008,
  month = oct,
  journal = {Reviews of Modern Physics},
  volume = {80},
  number = {4},
  pages = {1215--1274},
  issn = {0034-6861, 1539-0756},
  doi = {10.1103/RevModPhys.80.1215},
  urldate = {2023-09-29},
  langid = {english}
}

@article{golubchik2012,
  title = {Mass of a Vortex in a Superconducting Film Measured via Magneto-Optical Imaging plus Ultrafast Heating and Cooling},
  author = {Golubchik, Daniel and Polturak, Emil and Koren, Gad},
  year = 2012,
  month = feb,
  journal = {Physical Review B},
  volume = {85},
  number = {6},
  pages = {060504},
  publisher = {American Physical Society},
  doi = {10.1103/PhysRevB.85.060504},
  urldate = {2025-02-27}
}

@misc{grani2025,
  title = {Mutual friction and vortex Hall angle in a strongly interacting Fermi superfluid},
  author = {Grani, Nicola and Hernández-Rajkov, Diego and Daix, Cyprien and Pieri, Pierbiagio and Pini, Michele and Magierski, Piotr and Wlazłowski, Gabriel and Fernández, Marcia Frómeta and Scazza, Francesco and Pace, Giulia Del and Roati, Giacomo},
  year = 2025,
  month = mar,
  number = {arXiv:2503.21628},
  eprint = {2503.21628},
  primaryclass = {cond-mat},
  publisher = {arXiv},
  doi = {10.48550/arXiv.2503.21628},
  urldate = {2025-05-08},
  archiveprefix = {arXiv}
}

@article{hernandez-rajkov2024,
  title = {Connecting Shear-Flow and Vortex Array Instabilities in Annular Atomic Superfluids},
  author = {{Hernandez-Rajkov}, D. and Grani, N. and Scazza, F. and Pace, G. Del and Kwon, W. J. and Inguscio, M. and Xhani, K. and Fort, C. and Modugno, M. and Marino, F. and Roati, G.},
  year = 2024,
  month = jun,
  journal = {Nature Physics},
  volume = {20},
  number = {6},
  eprint = {2303.12631},
  primaryclass = {cond-mat},
  pages = {939--944},
  issn = {1745-2473, 1745-2481},
  doi = {10.1038/s41567-024-02466-4},
  urldate = {2025-03-20},
  archiveprefix = {arXiv}
}

@article{hu2006,
  title = {Equation of State of a Superfluid {{Fermi}} Gas in the {{BCS-BEC}} Crossover},
  author = {Hu, H. and Liu, X.-J. and Drummond, P. D.},
  year = 2006,
  month = apr,
  journal = {Europhysics Letters},
  volume = {74},
  number = {4},
  pages = {574},
  publisher = {IOP Publishing},
  issn = {0295-5075},
  doi = {10.1209/epl/i2006-10023-y},
  urldate = {2024-04-24},
  langid = {english}
}

@article{ichmoukhamedov2020,
  title = {Vortices in {{Fermi}} Gases with Spin-Dependent Rotation Potentials},
  author = {Ichmoukhamedov, T. and Tempere, J.},
  year = 2020,
  month = may,
  journal = {Physical Review A},
  volume = {101},
  number = {5},
  pages = {053609},
  issn = {2469-9926, 2469-9934},
  doi = {10.1103/PhysRevA.101.053609},
  urldate = {2025-03-03},
  langid = {english}
}

@article{kanjo2024,
  title = {Universal Description of Massive Point Vortices and Verification Methods of Vortex Inertia in Superfluids},
  author = {Kanjo, Akihiro and Takeuchi, Hiromitsu},
  year = 2024,
  month = dec,
  journal = {Physical Review A},
  volume = {110},
  number = {6},
  pages = {063311},
  publisher = {American Physical Society},
  doi = {10.1103/PhysRevA.110.063311},
  urldate = {2025-01-28}
}

@article{klimin2014,
  title = {Finite-Temperature Effective Field Theory for Dark Solitons in Superfluid {{Fermi}} Gases},
  author = {Klimin, S. N. and Tempere, J. and Devreese, J. T.},
  year = 2014,
  month = nov,
  journal = {Physical Review A},
  volume = {90},
  number = {5},
  pages = {053613},
  publisher = {American Physical Society},
  doi = {10.1103/PhysRevA.90.053613},
  urldate = {2022-08-04}
}

@article{klimin2015,
  title = {Finite Temperature Effective Field Theory and Two-Band Superfluidity in {{Fermi}} Gases},
  author = {Klimin, Serghei N. and Tempere, Jacques and Lombardi, Giovanni and Devreese, Jozef T.},
  year = 2015,
  month = may,
  journal = {The European Physical Journal B},
  volume = {88},
  number = {5},
  pages = {122},
  issn = {1434-6036},
  doi = {10.1140/epjb/e2015-60213-4},
  urldate = {2022-08-04},
  langid = {english}
}

@article{klimin2016,
  title = {Finite-Temperature Vortices in a Rotating {{Fermi}} Gas},
  author = {Klimin, S. N. and Tempere, J. and Verhelst, N. and Milo{\v s}evi{\'c}, M. V.},
  year = 2016,
  month = aug,
  journal = {Physical Review A},
  volume = {94},
  number = {2},
  pages = {023620},
  publisher = {American Physical Society},
  doi = {10.1103/PhysRevA.94.023620},
  urldate = {2024-01-11}
}

@article{kopnin1978,
  title = {Frequency Singularities of the Dissipation in the Mixed State of Pure Type-{{II}} Superconductors at Low Temperatures},
  author = {Kopnin, N B},
  year = 1978,
  journal = {JETP Letters},
  volume = {27},
  number = {7},
  pages = {390}
}

@article{kopnin1998,
  title = {Dynamic {{Vortex Mass}} in {{Clean Fermi Superfluids}} and {{Superconductors}}},
  author = {Kopnin, N. B. and Vinokur, V. M.},
  year = 1998,
  month = nov,
  journal = {Physical Review Letters},
  volume = {81},
  number = {18},
  pages = {3952--3955},
  issn = {0031-9007, 1079-7114},
  doi = {10.1103/PhysRevLett.81.3952},
  urldate = {2025-08-09},
  copyright = {http://link.aps.org/licenses/aps-default-license},
  langid = {english}
}

@article{ku2014,
  title = {Motion of a {{Solitonic Vortex}} in the {{BEC-BCS Crossover}}},
  author = {Ku, Mark J. H. and Ji, Wenjie and Mukherjee, Biswaroop and {Guardado-Sanchez}, Elmer and Cheuk, Lawrence W. and Yefsah, Tarik and Zwierlein, Martin W.},
  year = 2014,
  month = aug,
  journal = {Physical Review Letters},
  volume = {113},
  number = {6},
  pages = {065301},
  publisher = {American Physical Society},
  doi = {10.1103/PhysRevLett.113.065301},
  urldate = {2024-07-04}
}

@article{kwon2021,
  title = {Sound Emission and Annihilations in a Programmable Quantum Vortex Collider},
  author = {Kwon, W. J. and Del Pace, G. and Xhani, K. and Galantucci, L. and Muzi Falconi, A. and Inguscio, M. and Scazza, F. and Roati, G.},
  year = 2021,
  month = dec,
  journal = {Nature},
  volume = {600},
  number = {7887},
  pages = {64--69},
  publisher = {Nature Publishing Group},
  issn = {1476-4687},
  doi = {10.1038/s41586-021-04047-4},
  urldate = {2023-09-29},
  copyright = {2021 The Author(s), under exclusive licence to Springer Nature Limited},
  langid = {english}
}

@book{lamb1997,
  title = {Hydrodynamics},
  author = {Lamb, Horace},
  year = 1997,
  series = {Cambridge Mathematical Library},
  edition = {6. ed. repr. [1.Cambridge University Press paperback ed.]},
  publisher = {Cambridge University Press},
  address = {Cambridge},
  isbn = {978-0-521-05515-4 978-0-521-45868-9},
  langid = {english}
}

@book{lifshitz1980,
  title = {Statistical Physics},
  author = {Lifshitz, E. M. and Pitaevskii, L. P.},
  translator = {Sykes, John Bradbury and Kearsley, M. J.},
  year = 1980,
  series = {Course of Theoretical Physics},
  edition = {2nd ed. rev. and enl},
  number = {9},
  publisher = {Pergamon press},
  address = {Oxford},
  isbn = {978-0-08-023073-3},
  langid = {english},
  lccn = {530.13}
}

@article{lombardi2016,
  title = {Soliton-Core Filling in Superfluid {{Fermi}} Gases with Spin Imbalance},
  author = {Lombardi, G. and Van Alphen, W. and Klimin, S. N. and Tempere, J.},
  year = 2016,
  month = jan,
  journal = {Physical Review A},
  volume = {93},
  number = {1},
  pages = {013614},
  issn = {2469-9926, 2469-9934},
  doi = {10.1103/PhysRevA.93.013614},
  urldate = {2024-03-29},
  copyright = {http://link.aps.org/licenses/aps-default-license},
  langid = {english}
}

@phdthesis{lombardi2017,
  title = {Effective Field Theory for Superfluid {{Fermi}} Gases:  {{Application}} to Polarons and Solitons},
  author = {Lombardi, Giovanni},
  year = 2017,
  month = jun,
  urldate = {2023-10-06},
  school = {University of Antwerp}
}

@article{marini1998,
  title = {Evolution from {{BCS}} Superconductivity to {{Bose}} Condensation: Analytic Results for the Crossover in Three Dimensions},
  shorttitle = {Evolution from {{BCS}} Superconductivity to {{Bose}} Condensation},
  author = {Marini, M. and Pistolesi, F. and Strinati, G.C.},
  year = 1998,
  month = jan,
  journal = {The European Physical Journal B - Condensed Matter and Complex Systems},
  volume = {1},
  number = {2},
  pages = {151--159},
  issn = {1434-6036},
  doi = {10.1007/s100510050165},
  urldate = {2022-10-12},
  langid = {english}
}

@article{nakamura2024,
  title = {Picosecond Trajectory of Two-Dimensional Vortex Motion in ${\mathrm{FeSe}}_{0.5}{\mathrm{Te}}_{0.5}$ Visualized by Terahertz Second Harmonic Generation},
  author = {Nakamura, Sachiko and Matsumoto, Haruki and Ogawa, Hiroki and Kobayashi, Tomoki and Nabeshima, Fuyuki and Maeda, Atsutaka and Shimano, Ryo},
  year = 2024,
  month = jul,
  journal = {Physical Review Letters},
  volume = {133},
  number = {3},
  pages = {036004},
  publisher = {American Physical Society},
  doi = {10.1103/PhysRevLett.133.036004},
  urldate = {2025-02-27}
}

@article{navarro2013,
  title = {Dynamics of a {{Few Corotating Vortices}} in {{Bose-Einstein Condensates}}},
  author = {Navarro, R. and {Carretero-Gonz{\'a}lez}, R. and Torres, P. J. and Kevrekidis, P. G. and Frantzeskakis, D. J. and Ray, M. W. and Altunta{\c s}, E. and Hall, D. S.},
  year = 2013,
  month = may,
  journal = {Physical Review Letters},
  volume = {110},
  number = {22},
  pages = {225301},
  issn = {0031-9007, 1079-7114},
  doi = {10.1103/PhysRevLett.110.225301},
  urldate = {2025-02-27},
  copyright = {http://link.aps.org/licenses/aps-default-license},
  langid = {english}
}

@article{nozieres1985,
  title = {Bose Condensation in an Attractive Fermion Gas: {{From}} Weak to Strong Coupling Superconductivity},
  shorttitle = {Bose Condensation in an Attractive Fermion Gas},
  author = {Nozi{\`e}res, P. and {Schmitt-Rink}, S.},
  year = 1985,
  month = may,
  journal = {Journal of Low Temperature Physics},
  volume = {59},
  number = {3},
  pages = {195--211},
  issn = {1573-7357},
  doi = {10.1007/BF00683774},
  urldate = {2024-12-13},
  langid = {english}
}

@article{palestini2014,
  title = {Temperature Dependence of the Pair Coherence and Healing Lengths for a Fermionic Superfluid throughout the {{BCS-BEC}} Crossover},
  author = {Palestini, F. and Strinati, G. C.},
  year = 2014,
  month = jun,
  journal = {Physical Review B},
  volume = {89},
  number = {22},
  pages = {224508},
  publisher = {American Physical Society},
  doi = {10.1103/PhysRevB.89.224508},
  urldate = {2024-04-11}
}

@book{pethick2006,
  title = {Bose-{{Einstein}} Condensation in Dilute Gases},
  author = {Pethick, Christopher J. and Smith, Henrik},
  year = 2006,
  edition = {Reprinted with corrections},
  publisher = {Cambridge University Press},
  address = {Cambridge},
  isbn = {978-0-521-66194-2 978-0-521-66580-3},
  langid = {english}
}

@article{petrov2004,
  title = {Weakly {{Bound Dimers}} of {{Fermionic Atoms}}},
  author = {Petrov, D. S. and Salomon, C. and Shlyapnikov, G. V.},
  year = 2004,
  month = aug,
  journal = {Physical Review Letters},
  volume = {93},
  number = {9},
  pages = {090404},
  issn = {0031-9007, 1079-7114},
  doi = {10.1103/PhysRevLett.93.090404},
  urldate = {2023-09-29},
  langid = {english}
}

@article{pitaevskii1961,
  title = {Vortex {{Lines}} in an {{Imperfect Bose Gas}}},
  author = {Pitaevskii, L. P.},
  year = 1961,
  month = aug,
  journal = {Journal of Experimental and Theoretical Physics},
  volume = {13},
  pages = {451}
}

@book{pitaevskii2003,
  title = {Bose-{{Einstein}} Condensation},
  author = {Pitaevski{\u \i}, L. P. and Stringari, S.},
  year = 2003,
  series = {Oxford Science Publications},
  number = {116},
  publisher = {Clarendon Press},
  address = {Oxford},
  isbn = {978-0-19-850719-2},
  lccn = {QC175.47.B65 P58 2003}
}

@article{popov1973,
  title = {Quantum Vortices and Phase Transitions in {{Bose}} Systems},
  author = {Popov, V N},
  year = 1973,
  month = aug,
  journal = {Journal of Experimental and Theoretical Physics},
  volume = {37},
  number = {2},
  pages = {341--345},
  langid = {english}
}

@article{richaud2020,
  title = {Vortices with Massive Cores in a Binary Mixture of {{Bose-Einstein}} Condensates},
  author = {Richaud, Andrea and Penna, Vittorio and Mayol, Ricardo and Guilleumas, Montserrat},
  year = 2020,
  month = jan,
  journal = {Physical Review A},
  volume = {101},
  number = {1},
  pages = {013630},
  publisher = {American Physical Society},
  doi = {10.1103/PhysRevA.101.013630},
  urldate = {2025-01-15}
}

@article{richaud2021,
  title = {Dynamics of Massive Point Vortices in a Binary Mixture of {{Bose-Einstein}} Condensates},
  author = {Richaud, Andrea and Penna, Vittorio and Fetter, Alexander L.},
  year = 2021,
  month = feb,
  journal = {Physical Review A},
  volume = {103},
  number = {2},
  pages = {023311},
  issn = {2469-9926, 2469-9934},
  doi = {10.1103/PhysRevA.103.023311},
  urldate = {2024-01-25},
  langid = {english}
}

@misc{richaud2024,
  title = {Dynamical Signature of Vortex Mass in {{Fermi}} Superfluids},
  author = {Richaud, Andrea and Caldara, Matteo and Capone, Massimo and Massignan, Pietro and Wlaz{\l}owski, Gabriel},
  year = 2024,
  month = oct,
  number = {arXiv:2410.12417},
  eprint = {2410.12417},
  publisher = {arXiv},
  doi = {10.48550/arXiv.2410.12417},
  urldate = {2024-10-21},
  archiveprefix = {arXiv}
}

@book{saffman1993,
  title = {Vortex {{Dynamics}}},
  author = {Saffman, P. G.},
  year = 1993,
  month = jan,
  edition = {1},
  publisher = {Cambridge University Press},
  address = {Cambridge},
  doi = {10.1017/CBO9780511624063},
  urldate = {2025-01-23},
  copyright = {https://www.cambridge.org/core/terms},
  isbn = {978-0-521-47739-0 978-0-521-42058-7 978-0-511-62406-3},
  langid = {english}
}

@article{samson2016,
  title = {Deterministic Creation, Pinning, and Manipulation of Quantized Vortices in a {{Bose-Einstein}} Condensate},
  author = {Samson, E. C. and Wilson, K. E. and Newman, Z. L. and Anderson, B. P.},
  year = 2016,
  month = feb,
  journal = {Physical Review A},
  volume = {93},
  number = {2},
  pages = {023603},
  issn = {2469-9926, 2469-9934},
  doi = {10.1103/PhysRevA.93.023603},
  urldate = {2025-02-27},
  copyright = {http://link.aps.org/licenses/aps-default-license},
  langid = {english}
}

@article{simonucci2013,
  title = {Temperature Dependence of a Vortex in a Superfluid {{Fermi}} Gas},
  author = {Simonucci, S. and Pieri, P. and Strinati, G. C.},
  year = 2013,
  month = jun,
  journal = {Physical Review B},
  volume = {87},
  number = {21},
  pages = {214507},
  publisher = {American Physical Society},
  doi = {10.1103/PhysRevB.87.214507},
  urldate = {2024-04-11}
}

@article{simula2018,
  title = {Vortex Mass in a Superfluid},
  author = {Simula, Tapio},
  year = 2018,
  month = feb,
  journal = {Physical Review A},
  volume = {97},
  number = {2},
  pages = {023609},
  publisher = {American Physical Society},
  doi = {10.1103/PhysRevA.97.023609},
  urldate = {2024-12-04}
}

@article{suhl1965,
  title = {Inertial {{Mass}} of a {{Moving Fluxoid}}},
  author = {Suhl, H.},
  year = 1965,
  month = feb,
  journal = {Physical Review Letters},
  volume = {14},
  number = {7},
  pages = {226--229},
  publisher = {American Physical Society},
  doi = {10.1103/PhysRevLett.14.226},
  urldate = {2025-02-26}
}

@article{taylor2006,
  title = {Pairing Fluctuations and the Superfluid Density through the {{BCS-BEC}} Crossover},
  author = {Taylor, E. and Griffin, A. and Fukushima, N. and Ohashi, Y.},
  year = 2006,
  month = dec,
  journal = {Physical Review A},
  volume = {74},
  number = {6},
  pages = {063626},
  issn = {1050-2947, 1094-1622},
  doi = {10.1103/PhysRevA.74.063626},
  urldate = {2024-11-28},
  copyright = {http://link.aps.org/licenses/aps-default-license},
  langid = {english}
}

@article{tempere2009,
  title = {Effect of Population Imbalance on the {{Berezinskii-Kosterlitz-Thouless}} Phase Transition in a Superfluid {{Fermi}} Gas},
  author = {Tempere, J. and Klimin, S. N. and Devreese, J. T.},
  year = 2009,
  month = may,
  journal = {Physical Review A},
  volume = {79},
  number = {5},
  pages = {053637},
  issn = {1050-2947, 1094-1622},
  doi = {10.1103/PhysRevA.79.053637},
  urldate = {2023-10-11},
  langid = {english}
}

@incollection{tempere2012,
  title = {Path-{{Integral Description}} of {{Cooper Pairing}}},
  booktitle = {Superconductors - {{Materials}}, {{Properties}} and {{Applications}}},
  author = {Tempere, Jacques and Devreese, Jeroen P.A.},
  editor = {Gabovich, Alexander},
  year = 2012,
  month = oct,
  publisher = {InTech},
  address = {London},
  doi = {10.5772/48458},
  urldate = {2022-10-28},
  isbn = {978-953-51-0794-1},
  langid = {english}
}

@article{tesar2021,
  title = {Mass of Abrikosov vortex in high-temperature superconductor YBa$_2$Cu$_3$O$_{7-\delta }$},
  author = {Tesař, Roman and Šindler, Michal and Kadlec, Christelle and Lipavský, Pavel and Skrbek, Ladislav and Koláček, Jan},
  year = 2021,
  month = nov,
  journal = {Scientific Reports},
  volume = {11},
  number = {1},
  pages = {21708},
  issn = {2045-2322},
  doi = {10.1038/s41598-021-00846-x},
  urldate = {2025-02-27},
  langid = {english}
}

@article{vanalphen2024,
  title = {Splitting Instability of a Doubly Quantized Vortex in Superfluid {{Fermi}} Gases},
  author = {Van Alphen, W. and Takeuchi, H. and Tempere, J.},
  year = 2024,
  month = apr,
  journal = {Physical Review A},
  volume = {109},
  number = {4},
  pages = {043317},
  publisher = {American Physical Society},
  doi = {10.1103/PhysRevA.109.043317},
  urldate = {2025-01-10}
}

@article{vanloon2018,
  title = {Transition from Supersonic to Subsonic Waves in Superfluid {{Fermi}} Gases},
  author = {Van Loon, Senne and Van Alphen, Wout and Tempere, Jacques and Kurkjian, Hadrien},
  year = 2018,
  month = dec,
  journal = {Physical Review A},
  volume = {98},
  number = {6},
  pages = {063627},
  publisher = {American Physical Society},
  doi = {10.1103/PhysRevA.98.063627},
  urldate = {2022-08-04}
}

@article{verhelst2017,
  title = {Verification of an Analytic Fit for the Vortex Core Profile in Superfluid {{Fermi}} Gases},
  author = {Verhelst, Nick and Klimin, Serghei and Tempere, Jacques},
  year = 2017,
  month = feb,
  journal = {Physica C: Superconductivity and its Applications},
  volume = {533},
  pages = {96--100},
  issn = {09214534},
  doi = {10.1016/j.physc.2016.06.020},
  urldate = {2022-08-04},
  langid = {english}
}

@article{volovik1998,
  title = {Vortex Mass in {{BCS}} Systems: {{Kopnin}} and {{Baym-Chandler}} Contributions},
  shorttitle = {Vortex Mass in {{BCS}} Systems},
  author = {Volovik, G. E.},
  year = 1998,
  month = apr,
  journal = {Journal of Experimental and Theoretical Physics Letters},
  volume = {67},
  number = {7},
  pages = {528--532},
  issn = {1090-6487},
  doi = {10.1134/1.567692},
  urldate = {2024-11-14},
  langid = {english}
}

@article{wacker2015,
  title = {Tunable dual-species Bose-Einstein condensates of $^{39}\mathrm{K}$ and $^{87}\mathrm{Rb}$},
  author = {Wacker, L. and Jørgensen, N. B. and Birkmose, D. and Horchani, R. and Ertmer, W. and Klempt, C. and Winter, N. and Sherson, J. and Arlt, J. J.},
  year = 2015,
  month = nov,
  journal = {Physical Review A},
  volume = {92},
  number = {5},
  pages = {053602},
  publisher = {American Physical Society},
  doi = {10.1103/PhysRevA.92.053602},
  urldate = {2025-04-15}
}

@article{wilson2022,
  title = {Generation of High-Winding-Number Superfluid Circulation in {{Bose-Einstein}} Condensates},
  author = {Wilson, Kali E. and Samson, E. Carlo and Newman, Zachary L. and Anderson, Brian P.},
  year = 2022,
  month = sep,
  journal = {Physical Review A},
  volume = {106},
  number = {3},
  pages = {033319},
  publisher = {American Physical Society},
  doi = {10.1103/PhysRevA.106.033319},
  urldate = {2025-04-15}
}

@article{xu1995,
  title = {Ginzburg-{{Landau}} Equations for a {\emph{d}} -Wave Superconductor with Applications to Vortex Structure and Surface Problems},
  author = {Xu, Ji-Hai and Ren, Yong and Ting, C. S.},
  year = 1995,
  month = sep,
  journal = {Physical Review B},
  volume = {52},
  number = {10},
  pages = {7663--7674},
  issn = {0163-1829, 1095-3795},
  doi = {10.1103/PhysRevB.52.7663},
  urldate = {2025-03-05},
  copyright = {http://link.aps.org/licenses/aps-default-license},
  langid = {english}
}

@article{yefsah2013,
  title = {Heavy Solitons in a Fermionic Superfluid},
  author = {Yefsah, Tarik and Sommer, Ariel T. and Ku, Mark J. H. and Cheuk, Lawrence W. and Ji, Wenjie and Bakr, Waseem S. and Zwierlein, Martin W.},
  year = 2013,
  month = jul,
  journal = {Nature},
  volume = {499},
  number = {7459},
  pages = {426--430},
  publisher = {Nature Publishing Group},
  issn = {1476-4687},
  doi = {10.1038/nature12338},
  urldate = {2025-07-25},
  copyright = {2013 Springer Nature Limited},
  langid = {english}
}

@INPROCEEDINGS{wilson2024APS,
       author = {{Wilson}, Kali and {Moutamani}, Omar and {Despard}, Ilian},
        title = "{Vortex Dynamics in Binary Superfluids}",
    booktitle = {APS Division of Atomic, Molecular and Optical Physics Meeting Abstracts},
         year = 2024,
       series = {APS Meeting Abstracts},
       volume = {2024},
        month = jun,
          eid = {R09.007},
        pages = {R09.007},
       url = {https://ui.adsabs.harvard.edu/abs/2024APS..DMPR09007W}
}

\end{document}